\documentclass[11pt,a4paper]{article}

\pdfoutput=1
\usepackage{jcappub}
\usepackage{graphicx}
\usepackage{multirow}
\usepackage{hyperref}
\usepackage{amsmath,amssymb}
\usepackage{graphicx}
\usepackage{bm}
\usepackage{latexsym}
\usepackage{epsfig}
\usepackage{psfrag}
\usepackage{ulem}
\usepackage{color}
\usepackage[dvipsnames]{xcolor}
\usepackage{subfigure}
\usepackage[utf8]{inputenc}
\usepackage[a4paper, total={7in, 9in}]{geometry}
\makeatletter
\gdef\@fpheader{}
\makeatother
\newcommand{\be}{\begin{equation}}
\newcommand{\ee}{\end{equation}}
\newcommand{\bea}{\begin{eqnarray}}
\newcommand{\eea}{\end{eqnarray}}

\def\Mpl{M_{_{\mathrm{Pl}}}}

\allowdisplaybreaks

\begin{document}

\title{CMB Constraints on Natural Inflation with Gauge Field Production}
\author[a]{Khursid Alam,}

\author[a]{Koushik Dutta,}

\author[a,b]{and Nur Jaman}

\affiliation[a] {Department of Physical Sciences, Indian Institute of Science Education and Research Kolkata, Mohanpur-741 246, WB, India}
\affiliation[b]{{\color{blue}Department of Physics, Dhurba Chand Halder College, Dakshin Barasat,
South 24 Parganas,
Pin - 743372, West Bengal, India} }


\emailAdd{ka20rs114@iiserkol.ac.in}
\emailAdd{koushik@iiserkol.ac.in}
\emailAdd{nurphoton@gmail.com}






\abstract{
The natural inflation model with a periodic cosine potential is ruled out by recent Planck 2018 data for the decay constant $f \lesssim 5.5~M_{\rm Pl}$. If the Planck data is combined with the BICEP Keck array and BAO data, the model is excluded (at $2$-$\sigma$) for all values of $f$. In this context, we revisit the model when the pseudoscalar inflation $\phi$ is coupled with a gauge field via a coupling of the form $\frac{\alpha}{f} \phi F \tilde{F}$, where $F (\tilde F)$ denotes the gauge field (dual) strength tensor, and $\alpha$ is the coupling constant. The back-reactions associated with the gauge field production during the later stages of inflation extend the duration of inflation. We numerically evaluate the dynamics of the fields while neglecting the effects due to the perturbations in the inflaton field. It allows us to determine the scalar and tensor power spectra leading to the calculations of observables at the Cosmic Microwave Background (CMB) scales. We find that the natural inflation model survives the test of the latest data only for a certain range of the coupling constant $\alpha$. Our analysis shows that the latest constraints coming from the scalar spectral index are more stringent than the ones arising from the non-gaussianities and the running of the scalar spectrum. This leads to lower and upper bounds on $\xi_*$, the parameter that controls the growth of the gauge field.}

\maketitle

\section{Introduction}
Correlated superhorizon perturbations are well established from the Cosmic Microwave Background (CMB) radiation observations made by the Planck satellite \cite{Akrami:2018odb}. The rapid accelerated expansion of the early Universe or cosmic inflation can explain this feature of the perturbations at large scales \cite{Guth:1980zm,Starobinsky:1980te,Linde:1981mu,Albrecht:1982wi}. The idea of inflation also addresses the shortcomings of the hot big-bang model and explains the generation of primordial density fluctuations essential for large-scale structure formation. Various observations with high-precision data for temperature fluctuations and polarization measurements in the CMB have severely constrained inflationary models \cite{Martin:2013tda, Dutta:2021bih}.  

From the model-building perspective of inflation, one challenge is maintaining the required flatness of the potential against radiative corrections arising from several other degrees of freedom in the theory. One can ensure the required flatness of the potential if it is protected by some symmetry. The supersymmetry is not usually helpful as it is broken during inflation by the vacuum energy necessary for inflation to proceed \cite{Baumann:2014nda}. On the other hand, if a global symmetry is spontaneously broken at a scale $f$, the effective low energy theory enjoys a shift symmetry $\phi \rightarrow \phi + \text{constant}$, and $\phi$ becomes a Nambu-Goldstone-Boson. In the limit of exact shift symmetry, the potential is flat with the slow-roll parameters being zero. The shift symmetry gets mildly broken into a discrete subgroup by non-perturbative effects\footnote{The symmetry may be broken in several others ways leading to monomial potentials \cite{McAllister:2008hb, Kaloper:2008fb}.}. This leads to a periodic potential and the field becomes a pseudo-Nambu-Goldstone-Boson (pNGB) field $\phi$ responsible for driving inflation \cite{Freese:1990rb,Adams:1992bn}.
Due to the approximate shift symmetry, the pNGB potential is {\it natural} as it is stable against radiative corrections and provides a potentially UV complete theory of inflation. The model does not suffer the issues of perturbative radiative corrections since the periodic nature of the potential in natural inflation is generated from the non-perturbative sources. In the following discussions, we will occasionally call the pseud o-scalar an axion. 

The minimal model with a periodic potential for a pNGB \cite{Freese:1990rb,Adams:1992bn} of natural inflation is in tension with the recent data \cite{Akrami:2018odb}, especially for the sub-Planckian values of the symmetry-breaking scale $f \lesssim \Mpl$\footnote {Throughout the draft, we will be using the reduced Planck mass as $\Mpl$, and if not written explicitly, it will be set to $\Mpl =1$.}. The natural inflation model with a canonical kinetic energy term is ruled out (at $2$-$\sigma$) for all values of $f$ when the Planck data is combined with recent BICEP/Keck and BAO data; see Fig.~5 of reference \cite{BICEP:2021xfz}. The question that naturally arises is whether the model can be rescued by keeping its basic features intact. In this article, we would like to address this issue. The symmetry-breaking scale $f$ being larger than the Planck mass means that the global symmetry is broken above the quantum gravity scale where the validity of effective field theory is on shaky ground \cite{Arkani-Hamed:2003xts, Kallosh:1995hi,Salvio:2023cry}. At the same time, it has been conjectured that it is difficult to realize super-Planckian axion decay constant in any fundamental theory like in String Theory \cite{Banks:2003sx}. However, there exist models of many fields with sub-Planckian decay constants for an individual field, but effectively the models behave as one with an effective super-Planckian decay constant \cite{Kim:2004rp, Dimopoulos:2005ac, McAllister:2008hb, Kaloper:2008fb, Cicoli:2014sva, Das:2014gua}. Note that the CMB observables are sensitive only to this effective decay constant. The revised question that arises is that even for super-Planckian $f_{\rm eff}$, when is the model consistent with the data?

Other than the self-coupling, it is easily conceivable that the inflaton field is also coupled with the matter fields present in the theory. The pNGB field, due to the parity-violating nature (a pseudo-scalar), is generally coupled with the gauge field via the 5-dimensional operator $\mathcal{L}_{\rm int}  \supset - \frac{\alpha}{4\,f} \phi\,F\,\tilde{F}$, where $\alpha$ is the dimensionless coupling strength, $F ( \tilde F)$ is the gauge field (dual) tensors associated with some $U(1)$ gauge field $A_\mu$ and $f$ is the symmetry breaking scale. The coupling term allows inflaton energy to dissipate into the gauge fields and one of the helicity modes of the gauge field gets excited during inflation~\cite{Garretson:1992vt, Anber:2006xt,Ballardini:2019rqh,Campeti:2022acx}. This excited mode grows exponentially depending on the inflaton velocity dependent parameter $\xi (t) \equiv \frac{\alpha \, \dot{\phi_0}}{2 f H} $, where $\phi_0$ is the homogenous part of the inflaton field. As the velocity of the inflaton increases, the gauge fields are produced copiously and that starts to back react to the inflation dynamics leading to a prolonged duration of inflation \cite{Anber:2009ua, Barnaby:2011qe}. The production of gauge field also starts to affect the scalar perturbations via inverse decay of the gauge field, and the spectrum becomes highly non-gaussian at the CMB scales for $\xi_{*} \gtrsim 1$ \cite{Barnaby:2010vf, Barnaby:2011vw}. Throughout the paper, the quantities with the subscript $*$ indicate values at the CMB scale.

Due to the monotonic increase of $\xi$ as inflation progresses, the sourced contributions from the gauge field dominate the usual vacuum contributions and may produce primordial black holes at small scales~\cite{Linde:2012bt,Bugaev:2013fya, Cheng:2015oqa,Garcia-Bellido:2016dkw,Domcke:2017fix, Garcia-Bellido:2017aan,Cheng:2018yyr,Ozsoy:2023ryl,Cook:2022zol}.  The gauge quanta also serve as a source term for the parity-violating tensor power spectrum~\cite{Sorbo:2011rz}. Exponential dependence for the amplification of the gauge quanta on the parameter $\xi$ results in gravitational waves (GW) amplification on scales smaller than CMB~\cite{Sorbo:2011rz,Cook:2011hg,Barnaby:2011qe,Anber:2012du,Domcke:2016bkh,Garcia-Bellido:2016dkw,Garcia-Bellido:2023ser}. These high-frequency GW spectra may be probed by future missions~\cite{LISACosmologyWorkingGroup:2023njw} and will play a pivotal role in support of such models. Other phenomenological implications of this coupling have also been extensively studied which include magnetogenesis~\cite{Anber:2006xt,Caprini:2014mja,Adshead:2016iae} and baryogenesis \cite{Jimenez:2017cdr,Domcke:2019mnd,Domcke:2018eki,Talebian:2022cwk}.

The model can be considered in different regimes of the parameter $\xi(t)$ which controls the exponential growth of the gauge field.  During the evolution of the field, when the parameter $\xi$ becomes somewhat large, the amplified gauge fields significantly backreact to the background dynamics. The back-reaction occurs via an $\frac{\alpha}{f} \left\langle \bm{E} \cdot \bm{B} \right\rangle$ term in the equation of motion (EOM) of the inflaton and via the gauge-field energy density $\propto \left\langle \bm{E} \cdot \bm{E} \right\rangle + \left\langle \bm{B} \cdot \bm{B} \right\rangle$ in the Friedmann equation governing the Hubble rate. The back-reaction effects have been considered under different limits and approximations. For example, the authors of~\cite{Anber:2009ua} considered the evolution of inflaton field and its perturbations in a regime of strong back-reaction throughout the complete evolution of the field. In this case, the dissipation due to the gauge-field production provides the dominant source of friction for the inflaton motion; for recent analysis see \cite{vonEckardstein:2023gwk, Peloso:2022ovc}.  On the other hand, Ref.~\cite{Barnaby:2010vf, Barnaby:2011vw} considered the effects of the produced gauge field on inflaton perturbations at the CMB scales while neglecting its effect on the background evolution. However, this weak back-reaction regime at large scales quickly evolves to a strong back-reaction during the later stage of inflation \cite{Barnaby:2011qe}. It leads to the prolongation of inflation duration by about $10$ e-foldings. In this work, the authors have calculated the back-reaction terms due to gauge fields using the semi-analytical expression for $A_\mu$ evaluated under the adiabatic approximations.

In our work, we particularly focus on the effects of the gauge field on the inflationary dynamics and its imprints on CMB in light of the latest Planck, Keck, and BAO data \cite{Akrami:2018odb, BICEP:2021xfz}. Along with the Hubble parameter, we numerically solve the time evolution of the inflaton field and the mode equation for the gauge field. This allows us to calculate $\xi(t)$ even in the strong back-reaction regime. In the context of CMB observables, what is important is that due to the elongation of inflation when $\xi(t)$ becomes large, the pivot scale probes a different part of the inflation potential. Moreover, as we increase the coupling constant $\alpha$ for a fixed value of $f$, the behavior of the spectral index shows interesting features due to the interplay of contributions coming from vacuum fluctuations and the gauge field. It is important to note that we have calculated the scalar spectrum in terms of the background dynamics following the prescription of \cite{Linde:2012bt}. Similar numerical solutions, either in the present context or otherwise, of the system have been performed in the homogeneous limit of the inflation field~\cite{Cheng:2015oqa,Notari:2016npn,DallAgata:2019yrr,Sobol:2019xls,Cado:2022pxk,Domcke:2020zez}. In recent work, inhomogeneities of the inflation field have been included through lattice simulations \cite{Caravano:2022epk,Figueroa:2023oxc} and gradient expansion formalism \cite{Gorbar:2021rlt, vonEckardstein:2023gwk,Domcke:2023tnn} to study its effects at the small scales.

Our goal is to see whether the gauge field production in natural inflation with periodic potential can keep the model alive. The effects of gauge field can have broadly three regimes. Obviously, in one case, the effects can be so small (small $\alpha$) that it does not affect background and perturbations both at the CMB scales and small scales. 
In the second case, the production of gauge fields is very high to start with and therefore affects both background and perturbations throughout the evolution of the inflaton \cite{vonEckardstein:2023gwk, Peloso:2022ovc, Anber:2009ua,Durrer:2024ibi}.
Finally, the effects can be such that the gauge field affects the perturbations at the CMB scales, but not the background. But, due to the exponential growth of gauge fields, it affects the evolution at the later stages of inflation. We broadly work in this limit.  Therefore, in our case, the effects of back-reaction are small at $60$ e-foldings before the end of inflation, but we reach the strong back-reaction regime at a later stage. We will conclude that the model can be consistent only with the super-Planckian effective decay constant, and in that case also, only for a range of the coupling constant $\alpha$.

Our paper is organized in the following manner: In Sec.~\ref{sec:inflation}, we discuss inflationary dynamics driven by axion and its interaction with the gauge fields. We focus on calculating the back-reaction effects on the background variables, ignoring the effects generated by the inflaton fluctuations. Next, in Sec.~\ref{review:perturbations}, we discuss the basic setup for scalar and tensor perturbation for this scenario. In Sec.~\ref{sec:CMBprediction}, we show the effects of the gauge field on the CMB observables considering the numerical evaluation of the back-reaction done in the previous sections. The main result is summarised in the Fig~\ref{fig:N_s-vs-r.pdf}. Finally, we conclude our work in Sec.~\ref{Sec:Conclusion}, and also discuss the future outlook.

\medskip

\section{Pseudoscalar inflation with a gauge field}
\label{sec:inflation}


In this section, we discuss pseudoscalar inflaton $\phi$ coupled to an Abelian gauge field \cite{Anber:2006xt}. In particular, we will focus on the case of natural inflation where the inflaton potential is periodic~\cite{Freese:1990rb, Adams:1992bn}. In this case, a higher-dimensional coupling between the inflaton and the gauge field introduces an instability in the theory, and it leads to an exponential production of the gauge fields \cite{Garretson:1992vt,Anber:2006xt}. It is well known that this coupling induces a back-reaction both on the background dynamics~\cite{Anber:2009ua,Barnaby:2011qe} and in the perturbations~\cite{Anber:2012du, Linde:2012bt, Barnaby:2010vf, Barnaby:2011qe, Barnaby:2011vw}, leading to several observational consequences. 

Consider the action for a pseudo-scalar inflaton $\phi$, coupled to a massless Abelian gauge field $A_\mu$~\cite{Garretson:1992vt, Anber:2006xt, Anber:2009ua}:
\begin{equation}
\label{review:action_pseudoscalar}
\mathcal{S} = \int \textrm{d}^4 x \sqrt{-g} \left[ \frac{R}{2} - \frac{1}{2} \, \partial_\mu \phi \, \partial^\mu \phi - V(\phi) - \frac{1}{4} F_{\mu \nu} F^{\mu \nu} - \frac{\alpha}{4 f} \phi F_{\mu \nu} \tilde{F}^{\mu \nu} \right ]\ ,
\end{equation}
where $V(\phi) = \Lambda^4 \left[1 + \cos\left(\frac{\phi}{f}\right)\right]$ represents the inflaton potential for natural inflation~\cite{Freese:1990rb, Adams:1992bn}\footnote{Note that pseudo-scalar inflation can also be realized with several other forms of the $V(\phi)$ function, as discussed in~\cite{Barnaby:2011qe}.}. The inflationary scale is set by the constant $\Lambda$.  $F_{\mu \nu}$ and $\tilde{F}^{\mu \nu}$ denote the field-strength tensor and its dual respectively for the gauge fields, $f$ in the interaction term stands for a mass scale that suppresses the higher-dimensional scalar-vector coupling. The cosine potential arises from nonperturbative effects and breaks the continuous shift symmetry down to a discrete subgroup $\phi \rightarrow \phi + 2\pi f$. Note that the parameter $f$ that determines the periodicity of the inflation potential also sets the effective coupling between the gauge field and the inflation.  Additionally, $\alpha$ is the dimensionless coupling constant. 

It is worth noting that in the limit of $\alpha=0$, natural inflation is ruled out by the latest data \cite{Akrami:2018odb, BICEP:2021xfz}. For a fixed value of $f$, the height of the potential $\Lambda$ is determined by the value of the observed scalar power spectrum at the CMB scales. Once $f$ is fixed, the only remaining free parameter in the action of Eq.~\eqref{review:action_pseudoscalar} is $\alpha$. Thus, we can observe how the point in the $n_s$-$r$ plane changes by varying the coupling constant $\alpha$. For $\alpha \gg 1$, the effective mass scale $f/\alpha$ will take sub-Planckian values, as required for a reasonably effective field theory. 

We assume a homogeneous, isotropic, expanding universe described by a  Friedmann-Lema\^{i}tre-Robertson-Walker metric with a scale factor $a(t) \equiv a(\tau)$, where the cosmic time $t$ and conformal time $\tau$ are related by $d\tau = dt/a$.
 The Hubble rate and its conformal time counterpart are given by $H \equiv \dot{a}/a$ and $\mathcal{H} \equiv a'/a$, respectively, where dot corresponds to derivative w.r.t cosmic time and prime corresponds to derivative w.r.t conformal time. 
The dynamical equations of the inflation field, the Friedmann equation, and the gauge field are given respectively by \cite{Anber:2009ua}
\begin{align}
&\phi''+ 2\mathcal{H}\phi' -\mathbf{\nabla}^2\phi + a^2 \frac{\partial V}{\partial \phi}=a^2 \frac{\alpha}{f} \mathbf{E}\cdot \mathbf{B}~,
\label{EOMvarphi1}\\
&3\, \mathcal{H}^2= \left[\frac{1}{2}\phi'^2 + \frac{1}{2}(\mathbf{\nabla}\phi)^2 + a^2\, V(\phi)\, +\, \frac{a^2}{2}\left(\mathbf{E}^2 +  \mathbf{B^2}\right)\right], \label{Huble_equation}\\
& \mathbf{A}'' - \mathbf{\nabla}^2 \mathbf{A} + \mathbf{\nabla}(\mathbf{\nabla}.\mathbf{A}) = \frac{\alpha}{f}\phi' \left(\mathbf{\nabla}\times \mathbf{A}\right) - \frac{\alpha}{f}\left(\mathbf{\nabla}\phi\right)\times \mathbf{A}', \label{Maxwell_equations}\\
& (\mathbf{\nabla}. \mathbf{A})' = \frac{\alpha}{f}\left(\mathbf{\nabla}\phi\right).\left(\mathbf{\nabla}\times \mathbf{A}\right)~,
\end{align}
where we have used the temporal gauge, $A^{\mu} = (0, \mathbf{A})$. 
The above four equations are a complete set of coupled differential equations. Here we have introduced the physical electric and magnetic fields in terms of the gauge field
\begin{align}
    \mathbf{E} = -\frac{1}{a^2} \mathbf{A'}, \qquad \mathbf{B} = \frac{1}{a^2} \boldsymbol{\nabla} \times \mathbf{A}. \label{electromagnetic}
\end{align}

 \subsection{Gauge field production}
 \label{Gauge field production}

Let us briefly review gauge fields' production mechanism during inflation due to their coupling to the pseudoscalar inflaton ~\cite{Anber:2006xt, Anber:2009ua}. 
In the homogeneous limit we have $\mathbf{\nabla}\phi=0$, we can also consistently impose $\bm{\nabla}.\mathbf{A}=0$ for all time. In this case, the Eq.~ \eqref{Maxwell_equations} becomes, 
\begin{equation}
    \mathbf{A}'' - \mathbf{\nabla}^2 \mathbf{A} - \frac{\alpha}{f}\phi_{0}'\left(\mathbf{\nabla}\times \mathbf{A}\right)=0
    \label{gauge_fl_equ}.
\end{equation}
We decompose the above equation of motion of the gauge field in Fourier space. For this, we write down the gauge field in the momentum space,
\begin{equation}
\mathbf{A}(\mathbf{x}, \tau)  = \sum_{\lambda=\pm}\, \int \frac{d^3\,k}{(2\pi)^3}\, \left[A^{\lambda}_{\mathbf{k}}( \tau)\, \mathbf{\epsilon}_{\lambda}(\mathbf{\hat{k}}) \, a_{\lambda}(\mathbf{\hat{k}})\,e^{i\,\mathbf{k}.\mathbf{x}} + h.c \right] \label{decomposition_of_A},
\end{equation}
where $\pm$ represents the two helicities of the gauge field, $\mathbf{\epsilon}_{\lambda}(\mathbf{\hat{k}})$ represent the polarization vector of the gauge field which satisfies the condition $\mathbf{k}.\mathbf{\epsilon}^{\pm}(\mathbf{\hat{k}})=0, \mathbf{k} \times \mathbf{\epsilon}_{\pm}(\mathbf{\hat{k}}) = \mp i k\mathbf{\epsilon}_{\pm}(\mathbf{\hat{k}}), \mathbf{\epsilon}_{\pm}(-\mathbf{\hat{k}})= \mathbf{\epsilon}_{\pm}(\mathbf{\hat{k}})^{*}, \text{and}\,  \mathbf{\epsilon}_{\lambda}(\mathbf{\hat{k}})^{*}.\mathbf{\epsilon}_{\lambda'}(\mathbf{\hat{k}})= \delta_{\lambda\, \lambda'}$ and the annihilation  ($a_{\lambda}(\mathbf{k})$) and creation ($a^{\dagger}_{\lambda}(\mathbf{k})$) operators satisfy the commutation relation,
\begin{equation}
    [a_{\lambda}(\mathbf{k}),a^{\dagger}_{\lambda'}(\mathbf{k}')]= \delta_{\lambda\,\lambda'}\delta(\mathbf{k}-\mathbf{k}').
\end{equation}
After using the above conditions, we can write down the expression of the gauge field energy density and $\langle\mathbf{E}.\mathbf{B}\rangle$ in terms of gauge field modes $A_{k}^{\pm}$ as,
\begin{align}
&\frac{1}{2}\langle\mathbf{E}^2 + \mathbf{B}^2\rangle = \frac{1}{4\,\pi^2\, a^4}\int dk\, k^2\, \sum_{\lambda = \pm}\, \left(|A'^{\lambda}_{k}|^2 + k^2 |A^{\lambda}_{k}|^2\right) \label{energy_of_EM_fld},\\
&\langle \mathbf{E}. \mathbf{B}\rangle = -\frac{1}{4\,\pi^2\, a^4}\,\int\, dk\, k^3\, \frac{\partial}{\partial \tau}\left(|A^{+}_{k}|^2 - |A^{-}_{k}|^2\right), \label{edb_equation}
\end{align}
where $\langle..\rangle$ represent the averaging over space.

With the above mentioned mode expansion of Eq.~\eqref{decomposition_of_A}, we write down Eq.~\eqref{gauge_fl_equ} as,
\begin{equation}
    \left[\frac{\partial^2}{\partial \tau^2} + (k^2 \mp 2\,a\,H\,\xi\, k)\right]A^{\pm}_{k}(\tau)=0,\label{mode_of_gauge_fld}
\end{equation}
where $\xi = \frac{\alpha\, \dot{\phi_{0}}}{2\,f\,H}$.
From the above equation, we see that one of the two modes grows exponentially if that mode satisfies the condition $k/a\,H< 2|\xi|$. In another way, Eq.~\eqref{mode_of_gauge_fld} describes a tachyonic instability for the $A_{k}^{+}$ mode when we consider the initial velocity of the inflaton field to be positive. In this case, $\xi$ is always positive and increases on average till the end of inflation. During the initial stages of inflation, the value of $\xi$ remains nearly constant due to the slow-roll condition. Assuming $\xi$ to be constant, we can write down the solution of Eq. \eqref{mode_of_gauge_fld} for the modes, $(8\, \xi)^{-1} \lesssim k/(aH) \lesssim 2 \,\xi$. Additionally, we can also write down $\langle\mathbf{E}.\mathbf{B}\rangle$, and $\frac{1}{2}\langle\mathbf{E}^2+\mathbf{B}^2\rangle$ in terms of $A_{k}^{\pm}$ ~\cite{Anber:2009ua,Barnaby:2010vf,Barnaby:2011vw}:
\begin{align}
&A_{k}^{+} \simeq \frac{1}{\sqrt{2\, k}} \left( \frac{k}{2 \,  \xi \, a \, H}\right)^{1/4} e^{ \pi \xi - 2 \sqrt{2 \xi k/(a H)}}, \label{eq:Aprrox} \\
 &\langle \mathbf{E} \cdot \mathbf{B} \rangle \simeq  - \   2.4 \cdot 10^{-4} \frac{H^4}{\xi^4} e^{2 \pi \xi} \ , \quad \frac{1}{2} \langle \mathbf{E}^{2} + \mathbf{B}^{2} \rangle  \simeq \cdot \   1.4 \cdot 10^{-4} \frac{H^4}{\xi^3} e^{2 \pi \xi}.
  \label{eq:EdBapprox}
\end{align}
Note that the above expressions have exponential dependence on $\xi$, and as a result, their evolution also depends exponentially on it. 

As we are interested in studying the back-reaction effect of the gauge field on the background dynamics, we are neglecting the gradient of the inflation field. Hence, the background dynamics of the inflaton and the Hubble parameter are governed by the following equations
\begin{align}
    &\ddot{\phi}_0 + 3 H\dot{\phi}_0+\frac{dV}{d\phi_0} = \frac{\alpha}{f} \langle\mathbf{E}\cdot\mathbf{B}\rangle, \label{backr1}\\
    &3 H^2 = \frac{1}{2}\dot{\phi}_0^2  + V(\phi_{0}) + \frac{1}{2}\left\langle \mathbf{E}^2 + \mathbf{B}^2 \right\rangle~. \label{backr2} 
\end{align}
In our work, we solve the background dynamics of the inflaton field and the Hubble parameter Eqs.~\eqref{backr1} and \eqref{backr2} in the presence of the source term $\langle\mathbf{E}.\mathbf{B}\rangle$ and the term $\langle \mathbf{E}^2 + \mathbf{B}^2\rangle$. To calculate the source terms, we solve Eq.~\eqref{mode_of_gauge_fld} and evaluate $\langle\mathbf{E}.\mathbf{B}\rangle$, and $\langle \mathbf{E}^2 + \mathbf{B}^2\rangle$ by using the Eqs.~ \eqref{energy_of_EM_fld} and \eqref{edb_equation}. It is possible to demonstrate that the back-reaction term and the energy density of the gauge field in the Friedmann equation are relatively negligible for small values of $\xi$. However, during the later part of inflation, these terms cannot be neglected because $\xi$ starts to increase due to the rising velocity of the inflaton field. Effectively, these terms introduce an additional friction term to the inflaton dynamics that has an exponential dependence on $\xi$.
\subsection{Back-reaction of gauge fields on inflation dynamics }
\label{Back-reaction of gauge fields on inflation dynamics}
Let us now discuss the effects of the gauge field production on the inflationary background dynamics. The results of this subsection will be used subsequently to evaluate the scalar and tensor power spectrum. To understand the back-reaction effect of the gauge field on the inflationary dynamics, we solve the coupled differential equations \eqref{mode_of_gauge_fld}, \eqref{backr1}, and \eqref{backr2} simultaneously. In the Appendix \ref{appendix1}, we have provided a detailed explanation of the numerical procedure for solving the equations.

\begin{figure}[t]
\includegraphics[width=3.4in]{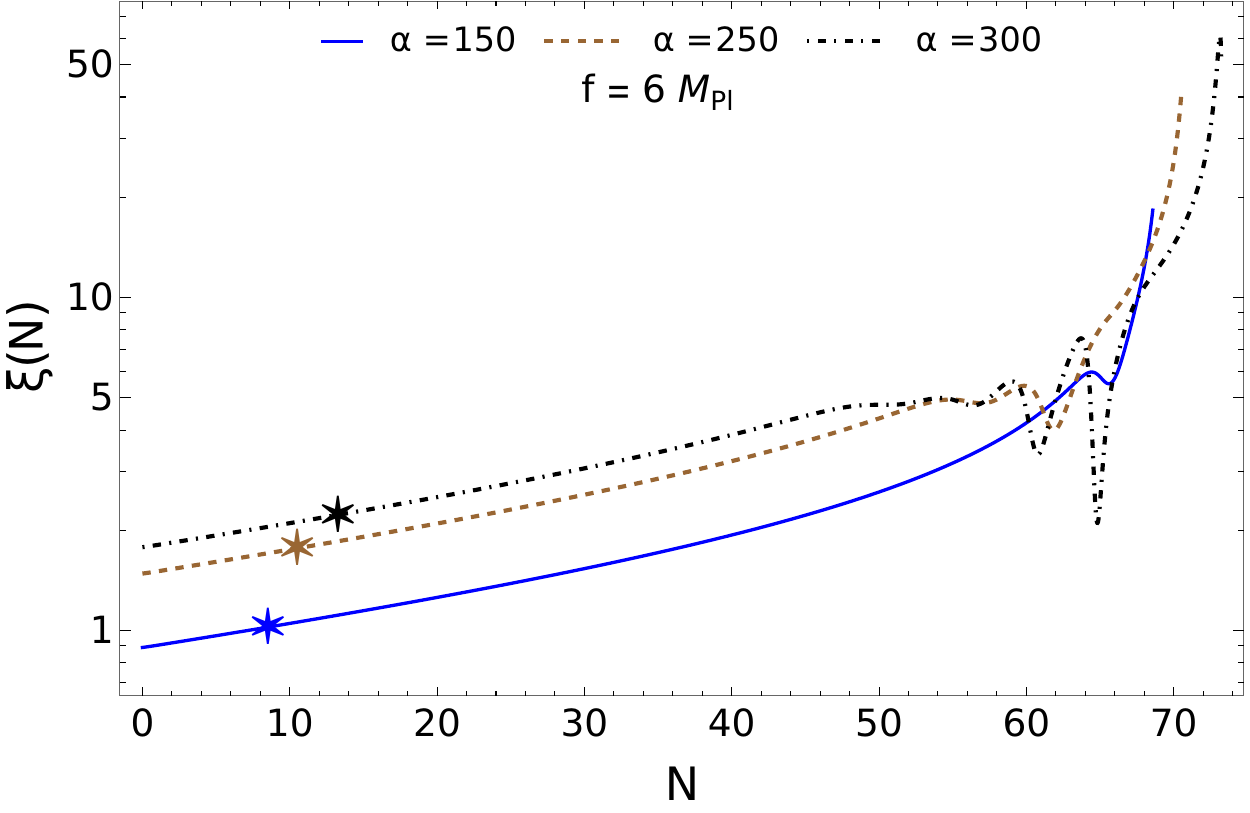}
\includegraphics[width=3.5in]{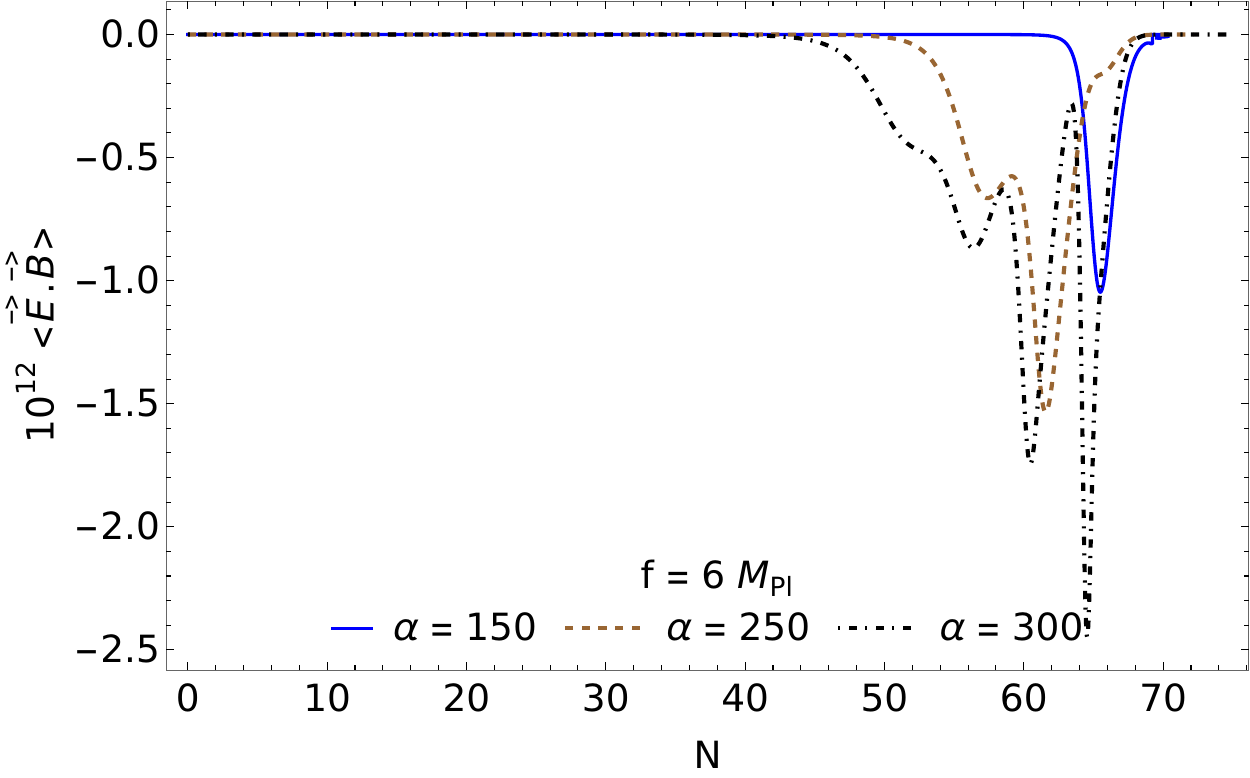}
\vskip -0.1in
\caption{The left panel depicts the time evolution of $\xi \equiv \frac{\alpha \dot{\phi_{0}}}{2fH}$ for various values of the coupling constant $\alpha$. The $\xi$ values are plotted until the end of inflation. The stars of different colors on the plot represent the points 60 $e$-foldings before the end of inflation. The right panel shows the time evolution of the back-reaction term (rescaled by $10^{12}$) for various values of the coupling constant $\alpha$. }
\label{fig:xi}
\end{figure}

The right-hand side of Eq.~\eqref{backr1}, i.e the term $\frac{\alpha}{f}\langle\mathbf{E}.\mathbf{B}\rangle$ is always negative (with our convention $A_{k}^{+}$ grows), which is shown in the Eqs.~\eqref{edb_equation}, and \eqref{eq:EdBapprox}. Therefore, the term acts as a source of damping for the background inflaton. The gauge field dynamics is governed by Eq.~\eqref{mode_of_gauge_fld}. It is evident  that this is mainly controlled by the term $\xi$ (see also Eq.~\eqref{eq:Aprrox}), which is directly proportional to the velocity of the inflaton. In the left panel of Fig.~\ref{fig:xi}, we show the evolution of $\xi$ for different values of the coupling constant $\alpha$. During the initial stages of inflation, the value of $\xi$ is small, typically being $\mathcal{O} (1)$. Hence, the less number of gauge field modes are excited as fewer gauge field quanta satisfy the condition $k< 2\,a\,H\, \xi$, which corresponds to the unstable solution. So, the production of gauge field is not significant at the initial stage of the evolution. As a result, the back-reaction term $\frac{\alpha}{f}\langle \mathbf{E} \cdot \mathbf{B}\rangle$ is negligible initially. In the right panel of Fig.~\ref{fig:xi}, we show the evolution of this term. Consequently, the back-reaction effect of the gauge field on the inflaton field dynamics is negligible. 

As inflation proceeds, the parameter $\xi$ grows because of increased field velocity. Consequently, more gauge field modes are excited, resulting in the back-reaction term becoming significant. This can be noticed in the right panel of Fig.~\ref{fig:xi} around the last $10$ - $20$ $e$-foldings of inflation. During this period, inflaton dynamics are affected by the produced gauge quanta, and it slows down the inflaton field, leading to reduced values of $\xi$. However, decreased values of $\xi$ correspond to a reduction in gauge field production because fewer modes are now being excited that satisfy the instability condition. This again lead to a decrease in back-reaction on the inflation dynamics, and inflaton velocity increases again. This tug of war between $\xi$ and back-reaction term ($\langle\mathbf{E}. \mathbf{B}\rangle$) may repeat depending on parameter choices. In several plots, we have shown this oscillatory behavior for several values of $\alpha$. These oscillations can be understood in terms of the memory effects of these nonlinear sets of equations \cite{Domcke:2020zez}.
\begin{figure}[t]
\includegraphics[width=3.57in]{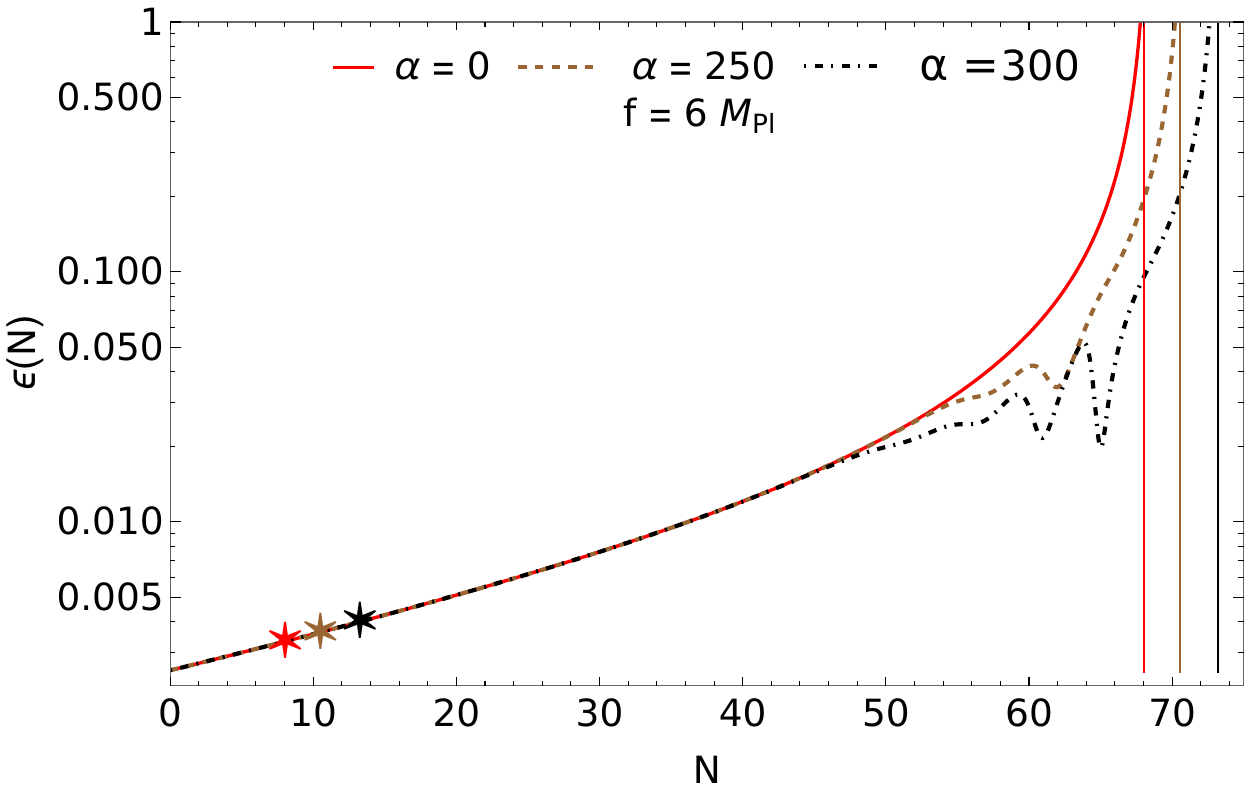}
\includegraphics[width=3.4in]{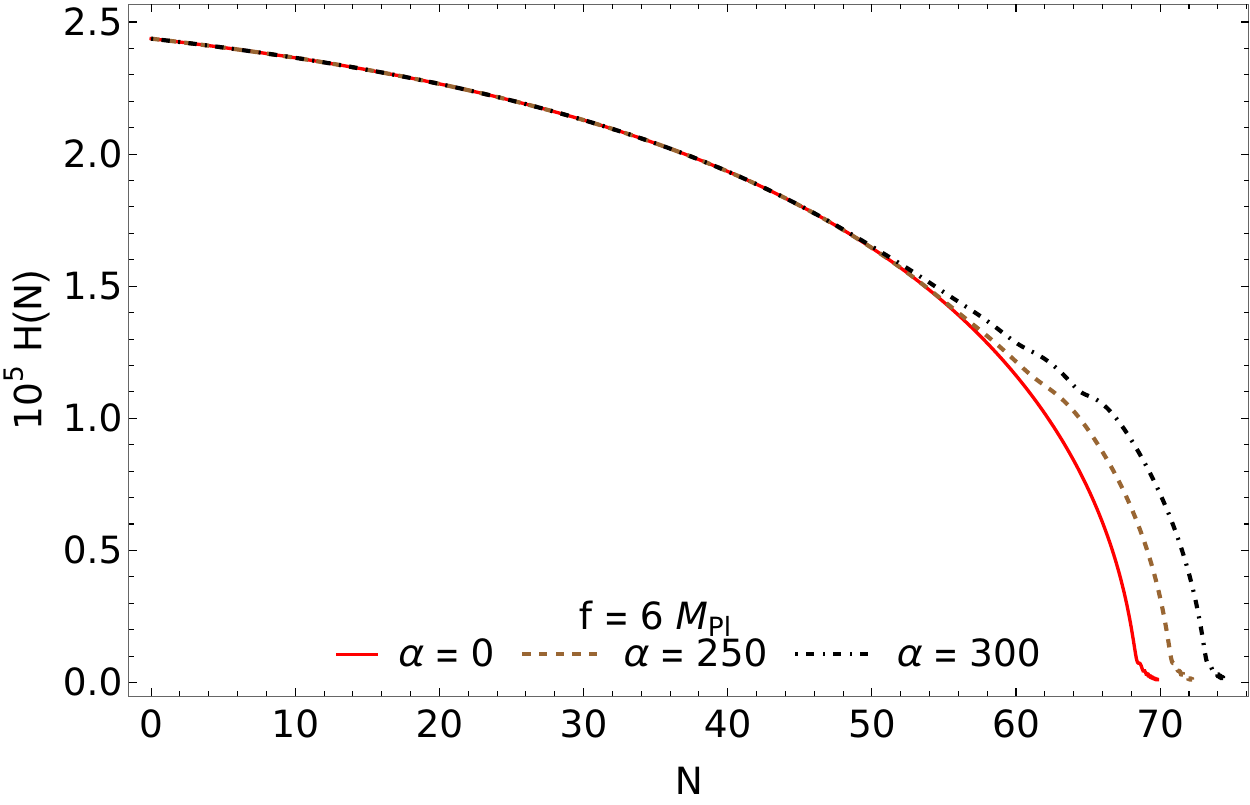}
\vskip -0.1in
\caption{The left and right panels of the figure show the time evolution of the first slow-roll parameter and the Hubble parameter for various values of the coupling constant $\alpha$. The vertical lines indicate the end of inflation ($\epsilon_H = 1$), while stars of different colors denote the 60 $e$ foldings before the end of inflation.}
\label{fig:epsilon1}
\end{figure}

\begin{figure}[t]
\includegraphics[width=3.5in]{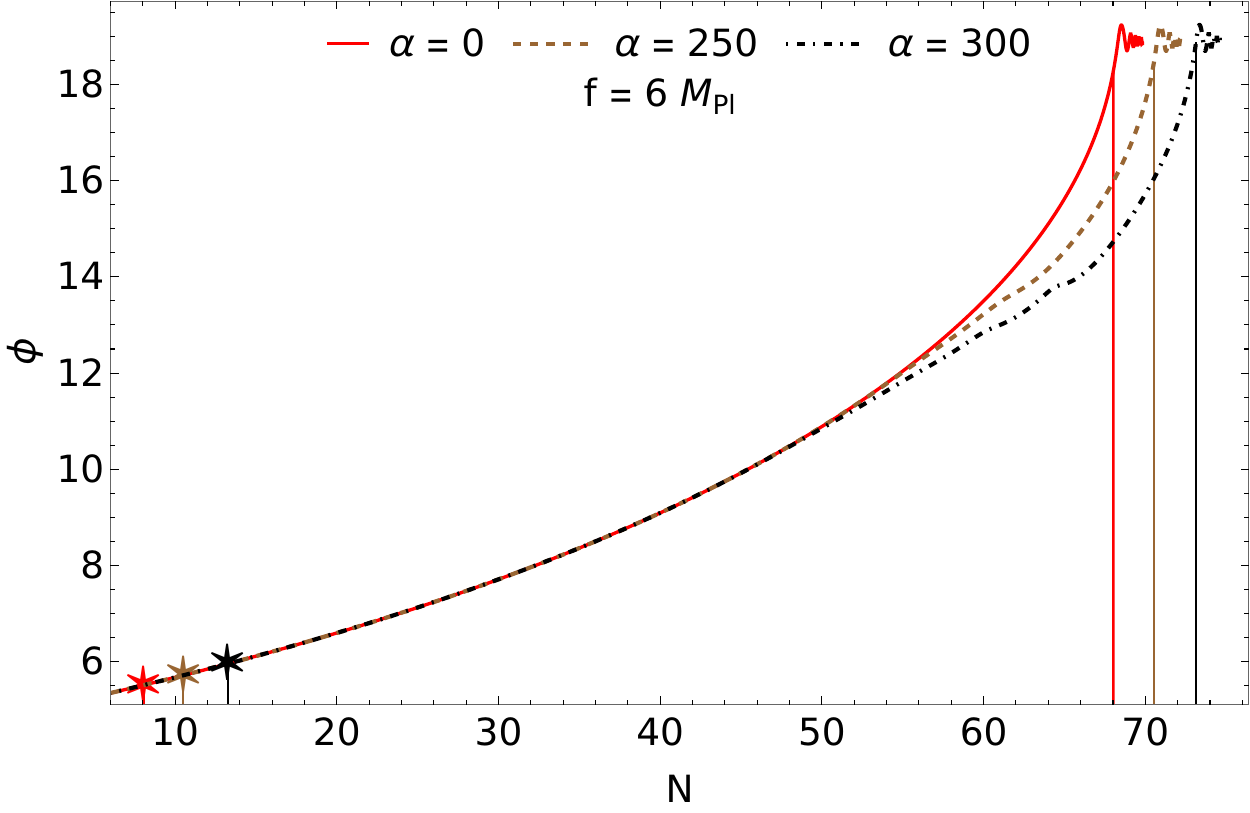}
\includegraphics[width=3.5in]{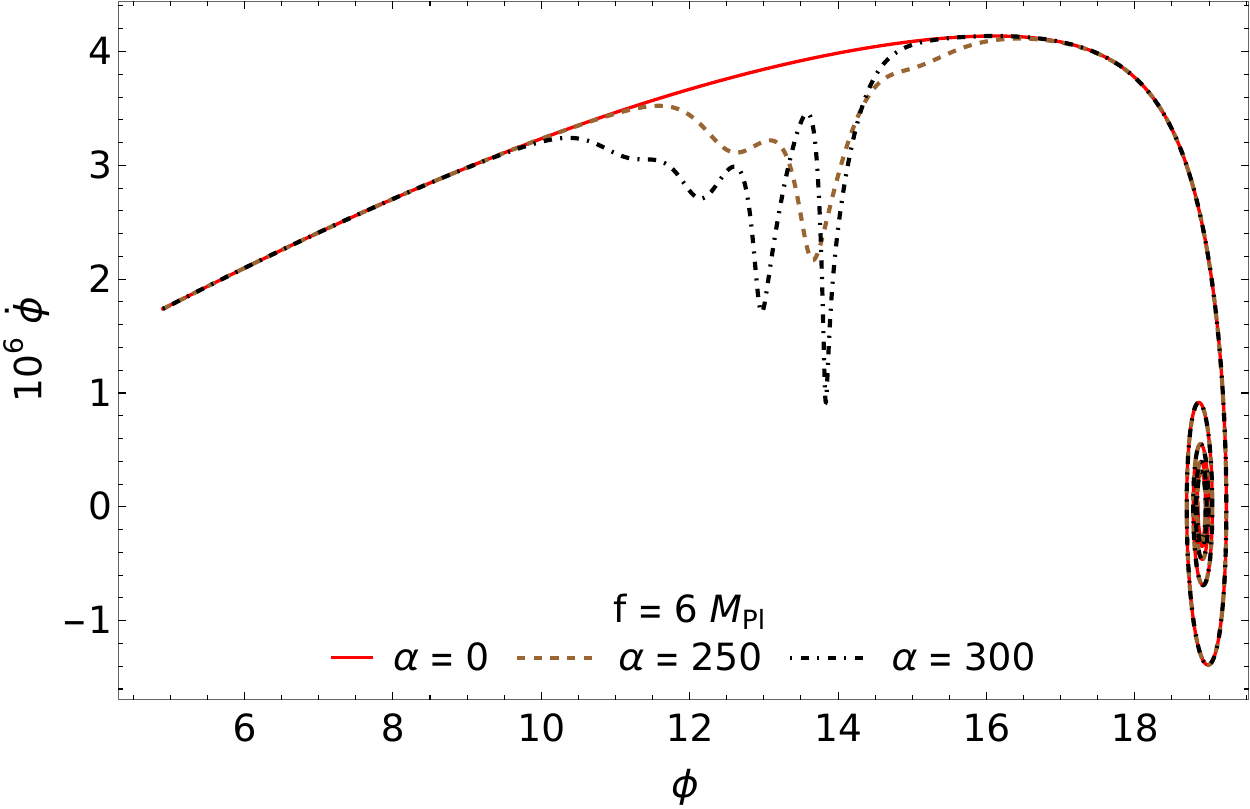}
\vskip -0.1in
\caption{Left panel is the time evaluation of the inflation field for different values of $\alpha$. The Stars and vertical lines represent the same things as the previous plots. The right panel is the phase space diagram of the inflaton field.}
{}
\label{fig:inflaton_dynmics}
\end{figure}

 In Fig.~\ref{fig:epsilon1}, the evolution of the first slow-roll parameter $\epsilon_{H}=-\dot{H}/H^2$ and the Hubble parameter are shown for various values of the coupling parameter $\alpha$. Inflation ends when $\epsilon_H = 1$ and these are marked by the vertical lines. Several stars correspond to the epoch of $60$ $e$-foldings before the end of inflation. From the $\epsilon_{H}$ plot, we observe that the presence of the gauge field prolongs inflation duration since the gauge field decelerate the dynamics of the inflaton field. In the right panel of Fig.~\ref{fig:epsilon1}, we have also depicted the evolution of the Hubble parameter with time. In Fig.~\ref{fig:inflaton_dynmics}, the inflaton dynamics and the phase diagram have been shown. From these figures, we can conclude that the initial inflaton dynamics mimic the usual cold inflationary behavior. \footnote{ The inclusion of gradient terms or spatial inhomogeneities in the inflaton field could affect the background dynamics, such as the duration of inflation, the Hubble parameter, and the slow-roll parameters, as shown in Fig.~\ref{fig:epsilon1} and Fig.~\ref{fig:inflaton_dynmics}. However, we believe that at the background level, these changes are not substantial enough to alter the overall scenario significantly (see \cite{Figueroa:2023oxc}).}

\section{Scalar and tensor power spectra}
\label{review:perturbations}
In this section, we briefly review the mechanism of producing scalar and tensor power spectrum that is affected by the produced gauge fields. In addition to the usual vacuum fluctuations, the inflation perturbations $\delta \phi = \phi - \phi_0(t)$ start to receive contributions from the process like $\delta {A} + \delta A\rightarrow \delta \phi$ as soon as the inflaton starts to roll \cite{Barnaby:2010vf, Barnaby:2011vw}. In this case, the fluctuations are governed by the following equation~\cite{ Barnaby:2011vw}
\begin{eqnarray}
    \delta \ddot\phi+ 3 H  \delta \dot\phi -\frac{\mathbf{\nabla}^2 \delta\phi }{a^2} + \frac{\partial^2 V}{\partial \phi_0^2} \delta \phi = \frac{\alpha}{f}\left( \mathbf{E}.\mathbf{B}- \langle \mathbf{E}.\mathbf{B}\rangle\right)\, , 
   \label{eq:pertScl1}
\end{eqnarray}
and at the very early stage of inflation when the effects of back-reaction are not so significant, the power spectrum can be approximately calculated as 
\begin{equation}
P_{\zeta}(k) = P_{\zeta}(k)_{\rm vac} (1 + P_{\zeta}(k)_{\rm vac} f_2 (\xi) e^{4\pi \xi} )~,
\label{scalar_amplitude}
\end{equation}
where $P_{\zeta}(k)_{\rm vac}$ is the vacuum contributions. The expression for $f_2(\xi)$ that is relevant for the CMB scales is given in \cite{Barnaby:2011vw}
\begin{equation}
    f_2(\xi) \simeq 3\cdot 10^{-5} \xi^{-5.4} ~~~~ 2 <\xi < 3~.
\end{equation}
The Eq.~\eqref{scalar_amplitude} is a good approximation for the power spectrum at the CMB scales, and it violates the scale invariance, as well as it is non-gaussian. This feature will be crucial for our later discussions. 

As inflation proceeds, the effects of produced gauge fields become substantial to affect the background dynamics of $\phi_{0}$ and $H$, as discussed above. This in turn affects the evolution of the perturbations as well. Moreover, as $\delta \phi$ grows, the source term in the r.h.s of Eq.~\eqref{eq:pertScl1} becomes $\delta \phi$ dependent. This effect can be estimated by modifying the perturbation equation with a modified friction term parameterized by $\beta$ \cite{Anber:2009ua, Barnaby:2011qe, Linde:2012bt},
\begin{eqnarray}
    {\delta  \ddot \phi}+ 3 H \beta  {\delta \dot \phi} -\frac{\mathbf{\nabla}^2 \delta\phi }{a^2} + V'' (\phi_{0})\delta \phi = \frac{\alpha}{f}\left( \mathbf{E}.\mathbf{B}- \langle \mathbf{E}.\mathbf{B}\rangle\right)\, , \label{eq:pertScl2}
\end{eqnarray}
where
\begin{eqnarray}
    \beta \equiv 1 - 2 \, \pi \, \xi \, \frac{\alpha}{f}\frac{\langle\mathbf{E}.\mathbf{B}\rangle}{3 H \dot \phi_{0}}\label{eq:beta}\, .
\end{eqnarray}
Note that $\beta$ is an indicator of whether the back-reaction at the perturbation level is important or not. This effect is negligible in the limit of $\beta \rightarrow 1$. 

The modified equation of motion has an approximated solution of the following form
\begin{eqnarray}
    \delta\phi \approx \frac{\alpha}{f} \frac{\left(\mathbf{E}.\mathbf{B}-\langle\mathbf{E}.\mathbf{B}\rangle\right)}{3 \beta H^2}~,
\end{eqnarray}
and it leads to the calculation of the scalar power spectrum valid even for large back-reaction 
\begin{eqnarray}
P_\zeta (k)_{\rm source} =  \frac{H^2}{\dot \phi_{0}^2} \langle\delta \phi^2\rangle =  \left(\frac{\alpha}{f}\frac{\langle\mathbf{E}.\mathbf{B}\rangle}{3\beta H\dot\phi_{0}}\right)^2\, . \label{pzeta_source}
\end{eqnarray}
It should be mentioned here that the above expression accounts for the contributions coming from the source term only. It should be noted that while evaluating the expression \eqref{pzeta_source}, spatial inhomogeneities were not taken into account. Including such terms could modify the behavior of the power spectrum at small scales. However, these small-scale modifications do not significantly impact the CMB scales. The total power spectra including the quantum vacuum fluctuations is 
\begin{eqnarray}
    P_\zeta (k) = P_\zeta(k)_{\rm vac}+ P_\zeta(k)_{\rm source}= \underbrace{ \left(\frac{H^2}{2\pi \dot \phi_{0}}\right)^2 }_{\rm{vaccum}}+ \,  \, \underbrace{\left(\frac{\alpha}{f}\frac{\langle\mathbf{E}.\mathbf{B}\rangle}{3\beta H\dot\phi_{0}}\right)^2}_{\rm sourced}~. \label{scalar_power_sp}
\end{eqnarray}
Thus the final power spectrum is twofold corrected. The first correction is due to the back-reaction effect in the background which we numerically solved. The second one is the back-reaction effect on the perturbations and it is parameterized by $\beta$. The above expression reduces to Eq.~\eqref{scalar_amplitude} for $\beta$ being close to $1$. We will use the above formula to calculate the scalar power spectrum for all length scales in Sec.~\ref{sec:CMBprediction}. Due to the introduction of the extra damping term via $\beta$, the spectrum at small scales goes as $\sim \xi^{-2}$, rendering the amplitude of the spectrum finite. 

Now, we discuss the production of gravitational waves due to the presence of the gauge fields during inflation. 
The equation of motion of the tensor fluctuation, $h_{ij}$ in the presence of gauge field is \cite{Barnaby:2011vw, Sorbo:2011rz}
\begin{equation}
    h''_{ij} + 2\frac{a'}{a}h'_{ij} - \nabla^2\,h_{ij} = \frac{2}{M^2_{\rm pl}} \Pi^{ lm}_{ij} T^{\rm EM}_{lm}. \label{EOM_tensor_mode}
\end{equation}
Here, $\Pi_{ij}^{lm} = \Pi_{i}^{l} \Pi_{j}^{m} - \frac{1}{2}\Pi_{ij}\Pi^{lm}$ represents the transverse and traceless projection operator, where $\Pi_{ij}= \delta_{ij} - \frac{\partial_{i}\partial_{j}}{\nabla^2}$, and $T_{ij}^{\text{EM}}$ denotes the spatial component of the total energy-momentum tensor of the gauge field, as discussed in \cite{Barnaby:2011vw, Sorbo:2011rz}. The energy-momentum tensor can be obtained by varying the action of Eq.~\eqref{review:action_pseudoscalar} with respect to the metric.

Solving Eq.~\eqref{EOM_tensor_mode} using the Green function method,
the expression for the tensor power spectrum in the presence of gauge field is \cite{Sorbo:2011rz,Barnaby:2011vw, Barnaby:2011qe}
\begin{equation}
    P_{h}^{\pm}= \frac{H^2}{\pi^2}\left(\frac{k}{k_{0}}\right)^{n_T}\left[1+ 2\,H^2 f^{\pm}_{h}(\xi) e^{4 \, \pi \, \xi}\right] \label{gw_spectrum},
\end{equation}
where $n_{T} = 2\epsilon_{H}$, $\epsilon_{H}$ is the first slow-roll parameter, and $\pm$ represents two polarization of the gravitational waves, where$f_{h}^{\pm}(\xi)$ are given by
\begin{equation}
    f^{+}_{h}= \frac{4.3\times 10^{-7}}{\xi^6}; ~~ f^{-}_{h}= \frac{9.2\times 10^{-10}}{\xi^{6}}; ~~ \text{where}~~ \xi \gg 1. 
\end{equation}
So, the total tensor power spectra is,
\begin{equation}
P_{h}= P^{+}_{h} + P^{-}_{h}\label{total_power_sp},
\end{equation}

In Eq.~\eqref{gw_spectrum}, we have written down the tensor power spectra in terms of the background quantities. So, we can calculate the tensor power spectrum after plunging the background evaluation of the Hubble parameter and $\xi$, and we have calculated those quantities in the sec \ref{Back-reaction of gauge fields on inflation dynamics}. In the last stage of inflation, significant production of gauge field will change the above-mentioned perturbation equation. As opposed to the scalar spectrum via the introduction of $\beta$, this effect has not been taken into account for the tensor modes. 
So, the tensor power can diverge (spuriously) at the last stage of inflation.

Even though the gravitational waves generated from the tensor power spectrum have not been detected at the CMB scales, there is a strict constraint on the ratio of the tensor power spectrum to the scalar power spectrum. This parameter, dubbed as tensor-to-scalar ratio, is defined as
\begin{align}
    r \equiv \frac{P_h}{P_\zeta} = \frac{P_h^+ + P_h^-}{P_\zeta}. \label{r}
\end{align}
Combining the latest Planck 2018 data with the BICEP-Keck data constrains this at $r_* \leq 0.056$ at the CMB scales \cite{Akrami:2018odb}.

\section{Imprints on CMB observables}
\label{sec:CMBprediction}
In sec \ref{sec:inflation}, we have discussed the dynamics of the background inflaton field as well as the behavior of the Hubble parameter, and in sec \ref{review:perturbations} we have reviewed the necessary formulas for the scalar and tensor power spectrum.  Armed with these details, we turn our attention to the calculations of the scalar and tensor power spectrum produced during inflation in the presence of a gauge field. It will lead us to calculate several observables that can be confronted with the latest data. Our goal is to find parameter values $\alpha$ and $f$ for which the model defined in Eq.~\eqref{review:action_pseudoscalar} is consistent with all existing observational data.  

To account for the uncertainty associated with the entropy generation at the end of inflation, the inflationary observables are typically calculated at $N_* = 50$ or $60$, where $N_*$ is the number of e-foldings back from the end of inflation to the time when CMB modes exited the horizon. The CMB pivot scale is at comoving wave number $k_* = 0.05$ Mpc$^{-1}$. Initially, we will discuss the results for the case $N_* = 60$, and later we will mention the results for $N_* = 50$. Because of rapid gauge field productions at the later stages of inflation,  the back-reaction on the inflaton dynamics becomes important and inflaton slows down for a while, and it allows the inflation duration to be prolonged. It is evident from the left panel of Fig.~\ref{fig:inflaton_dynmics}, where the stars (`$\star$') of different colours correspond to 60 e-foldings before the end of inflation for different choices of $\alpha$. We see that for different choices of parameter $\alpha$ (for a given value of $f$), the CMB scales probe different parts of the axion potential. As noted in Eqs. \eqref{scalar_power_sp} and \eqref{total_power_sp}, the fluctuations are governed by both the vacuum part as well as by the gauge quanta productions. The phenomenology of the interplay of these two contributions is interesting, and that will be discussed now.

\begin{figure}[t]
\includegraphics[width=3.5in]{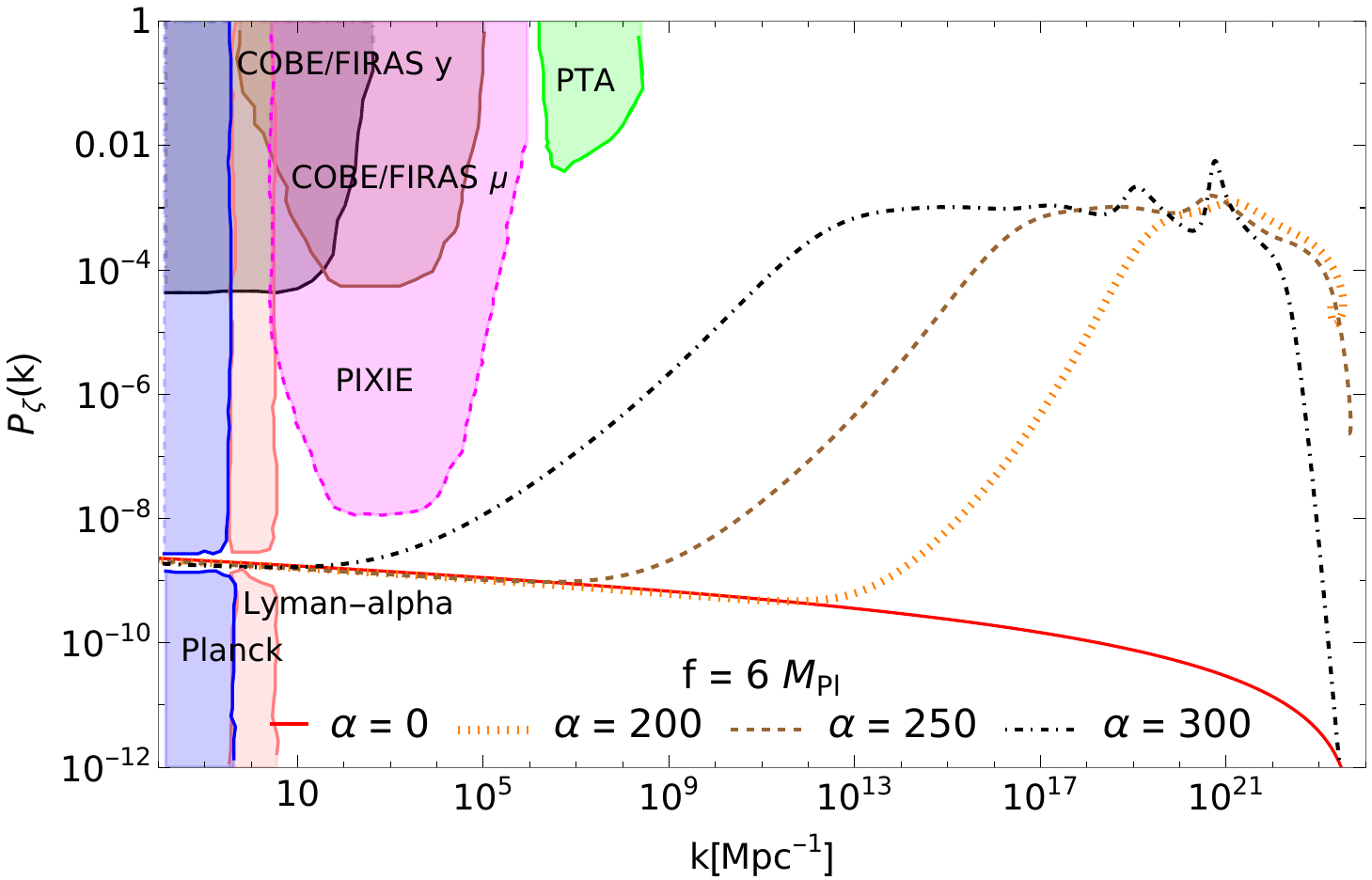}
\includegraphics[width=3.5in]{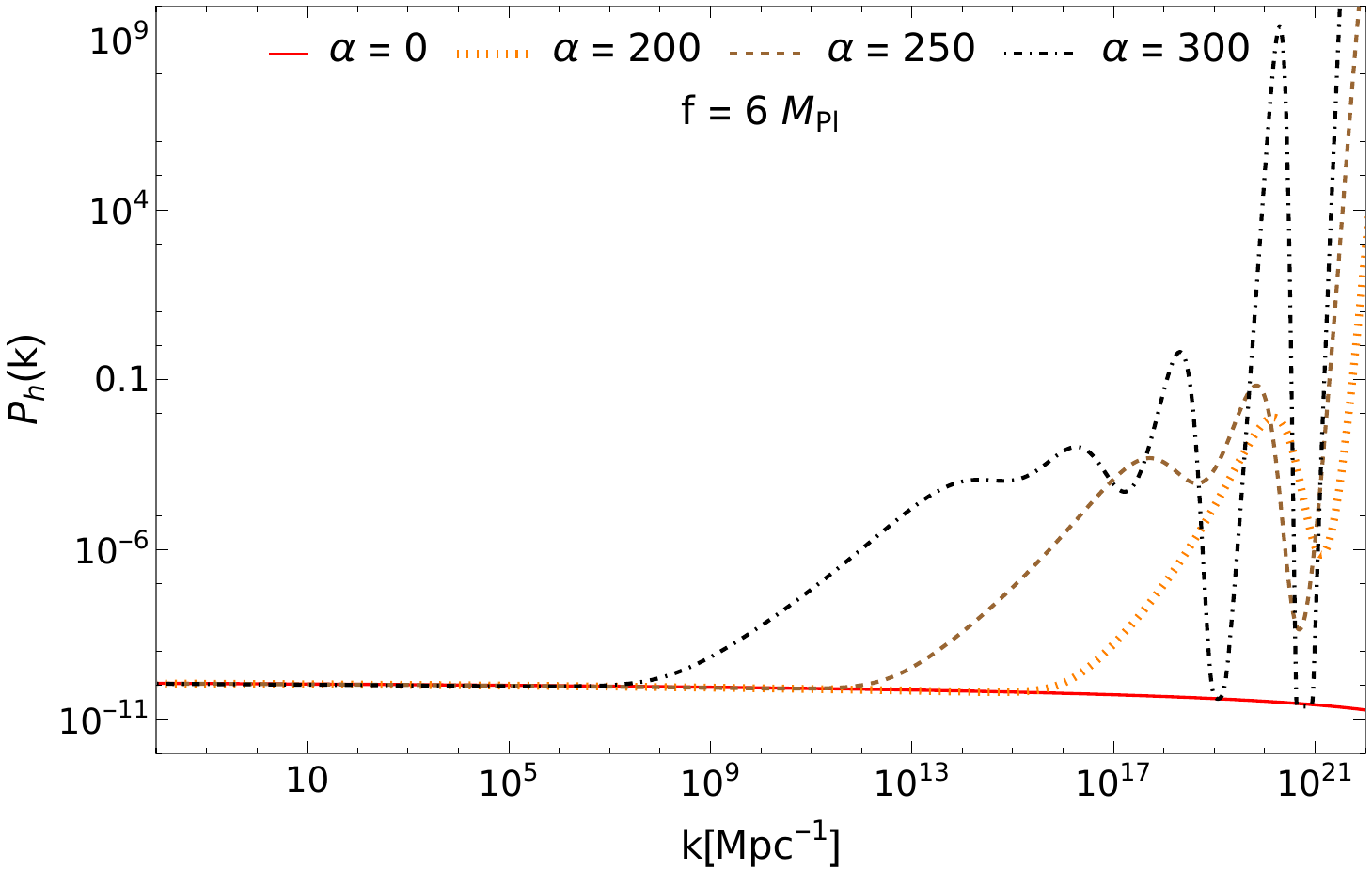}
\vskip -0.1in
\caption{In the left panel, the amplitude of the scalar power spectrum is plotted against comoving wavenumbers for different values of the coupling constant $\alpha$ and $f = 6$. Several colored contours show existing (continuous) and projected future (dotted) constraints. In the right panel, the amplitude of the tensor power spectrum is plotted against comoving wave numbers for the same values of the parameters as the scalar spectrum.}
\label{fig:scalar_tensor_power_spectra}
\end{figure}

In practice, we evaluate the scalar power spectrum as a function of the number of e-foldings $N$ using the values of background quantities appearing in the  Eq. \eqref{scalar_power_sp}. Note that the effects of back-reaction are incorporated in the background quantities. However, as previously mentioned, the spatial inhomogeneities of the inflaton field have not been taken into account.  We subsequently convert the spectrum in terms of comoving momenta $k$ to produce the left panel of Fig.~\ref{fig:scalar_tensor_power_spectra} where scalar spectrum $P_{\zeta}$ is plotted against $k$. In the plot, we show the scalar power spectrum for different values of $\alpha$ for the fixed value of $f = 6$. The red line corresponds to the usual slow-roll inflation in the absence of a gauge field. 
The scalar power spectrum shows several interesting features for $\alpha \neq 0$. The most prominent one is the sharp rise in the value of the spectrum at small scales. The rise of the spectrum happens for smaller $k$ values (larger wavelength) for larger $\alpha$.  Also, due to the strong back-reaction at the later stage of inflation, the power spectrum shows oscillatory features at small scales. It is essential to reemphasize that the spectrum does not diverge at smaller scales due to the consideration of the back-reaction effect in calculations of both the scalar power spectrum and $\xi$.  In this section, we focus on the power spectrum at the CMB scales.
In the plot, we also show several existing and future observational constraints on the scalar power spectrum from Planck \cite{Akrami:2018odb}, Lyman-alpha \cite{Bird_2011}, PIXIE \cite{A_Kogut_2011}, COBE/Firas \cite{Fixsen:1996nj}, and  PTA \cite{Byrnes:2018txb}. 

\begin{figure}[h]
\centering
\includegraphics[width=5.5in, height = 3 in]{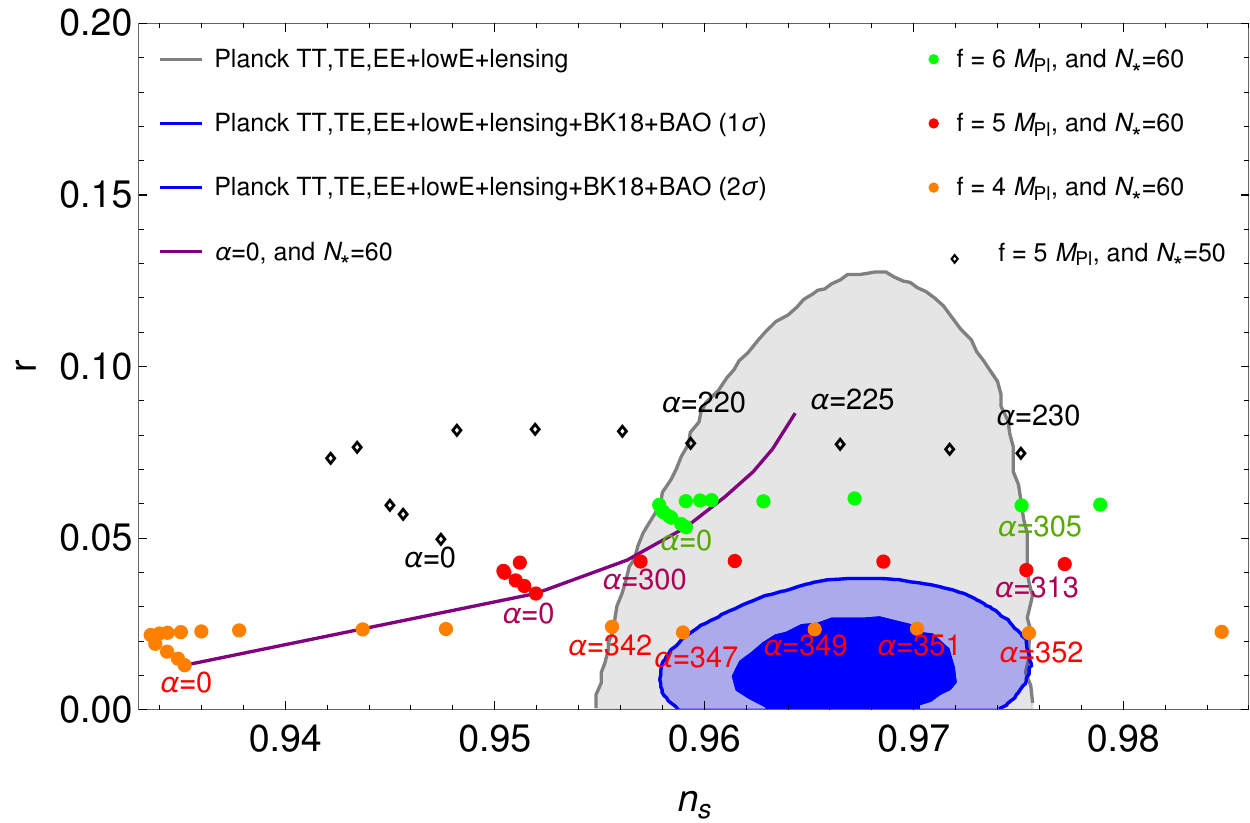}
\vskip -0.1in
    \caption{The scalar spectral index $n_s$ vs the tensor-to-scalar ratio $r$ for various values of the coupling constant $\alpha$. The continuous purple line corresponds to the case for $\alpha =0$ as we vary the value of $f$. The orange, red, and green dots of different colors correspond to different values of $f$ for $N_* = 60$. On the other hand, black diamond corresponds to $N_* = 50$ and $f = 5$. The  grey colored contour shows the $2$-$\sigma$ constraints coming from only Planck18 data, and the blue  contours  show the $1$-$\sigma$ (darker) and $2$-$\sigma$ (lighter) constraints when the Planck18 data \cite{Akrami:2018odb} is combined with BK18 and BAO \cite{BICEP:2021xfz}.}
\label{fig:N_s-vs-r.pdf}
\end{figure}

Using the formula of Eq.~\eqref{total_power_sp}, the same procedure is followed to evaluate the tensor power spectrum, and it is shown in the right panel of Fig. \ref{fig:scalar_tensor_power_spectra}. In contrast to the scalar power spectrum plot, the amplitude of the tensor power spectrum diverges at small scales. In this case, the regulated formula for the tensor power spectrum, accounting for the back-reaction, is unknown. Nevertheless, our focus is on the CMB physics (small $k$), and the back-reaction effect on the tensor amplitude is negligible around the CMB scale. So, we can utilize this formula to calculate the tensor-to-scalar ratio at the CMB scales. Note that for the case of scalar amplitude, the relevant observable is the slope of the spectrum ($n_s$), whereas for the tensor spectrum, the observable is related to the value of the amplitude itself.  
In the right panel of Fig.~\ref{fig:scalar_tensor_power_spectra}, the solid red represents the tensor power spectrum in the absence of the gauge field. The various other lines represent the tensor power spectrum in the presence of the gauge field, corresponding to different values of the coupling constant $\alpha$ for $f = 6$.  

Now, we proceed to calculate inflationary observables, namely scalar spectral index $n_s$ and tensor to scalar ratio $r$. This can be done straight away by numerically evaluating the derivative of the power spectrum at the pivot scale
\begin{equation}\label{eq:ns}
	n_s-1 \equiv \left.\frac{d \ln P_\zeta (k)}{d\ln k} \right\vert_{k_*}\, , 
\end{equation}
and taking the ratio $P_\zeta(k)/P_h(k)$ at the pivot scale. In Fig. \ref{fig:N_s-vs-r.pdf}, we show the model predictions for $n_s$ and $r$ as we vary $\alpha$ for several values of $f$. We also show $2$-$\sigma$ allowed contours for 2018 Planck TT, TE, EE+lowE+lensing \cite{Akrami:2018odb} (henceforth will be mentioned as Planck18) and $1$-$\sigma$ and $2$-$\sigma$ contours for 2018 Planck TT, TE, EE+lowE+lensing+BK18+BAO \cite{BICEP:2021xfz} (henceforth will be called Planck18+BK18). The solid purple line in Fig.~\ref{fig:N_s-vs-r.pdf} shows the model predictions $(N_* = 60)$ for canonical pNGB potential when $f$ is continuously varied. First of all, the model is consistent at $2$-$\sigma$  for $f \gtrsim 5.5$ only when the Planck18 data is considered. On the other hand, the model is excluded at $2$-$\sigma$ for all values of $f$ for Planck18+BK18 data. In this sense, the pNGB inflation model with periodic potential is under tension with the latest observational data. Note that reducing the value of $N_*$ makes the situation worse.

To understand the effects of gauge field production, we first focus on the interesting case with $f = 4$ and $N_* = 60$, as we vary $\alpha$. These variations are shown by the orange dotted points in Fig.~\ref{fig:N_s-vs-r.pdf} where $\alpha = 0$ is well outside the Planck18 $2$-$\sigma$ contour. It is noted above, for $\alpha = 0$, the model is excluded by Planck18 data at $2$-$\sigma$. As we increase the value of the coupling constant $\alpha$ from zero, the spectral index $n_s$ initially decreases, moving further away from the 2-$\sigma$ contour provided by the Planck18 data. Interestingly, the value of the spectral index makes a turn and then it starts to increase as the value of $\alpha$ is raised. At some point, it becomes consistent with the latest Planck18 data. If we continue increasing $\alpha$, the predictions eventually enter $2$-$\sigma$ followed by $1$-$\sigma$ contours even for the combined data set Planck18 + BK18. In this case, the spectral index remains consistent with the combined data set for $347\lesssim \alpha\lesssim 352$ at the $2$-$\sigma$ level and for $349\lesssim \alpha\lesssim 351$ at the $1$-$\sigma$ level (see the orange dots in Fig.\ref{fig:N_s-vs-r.pdf}). Therefore, a coupling between the gauge field and the pseudo-scalar inflaton helps to alleviate the tension of the pNGB model and makes it consistent in the $n_{s}$-$r$ plane. In summary, natural inflation with $\alpha = 0$ is very strongly disfavored by Planck18+BK18 data, irrespective
of the value of $f$. As an example, we see that for $f = 4$, a narrow range of $\alpha$  brings the model to agree with Planck+Keck.

The above-mentioned trend of $n_s$ dependence on $\alpha$ remains valid for higher values of $f$- see red and green dots of $f = 5$ and $f = 6$ respectively. In the case of $f = 5$, the model becomes consistent ($2$-$\sigma$) with only Planck18 data for some range of $\alpha$ values, but it never becomes consistent with Planck18+BK18 data at $2$-$\sigma$. For the case of $f = 6$, the model remains consistent ($2$-$\sigma$) with Planck18 data for $\alpha \lesssim 305$.

\begin{figure}[h]
\centering
\includegraphics[width=4.5 in, height = 3 in]{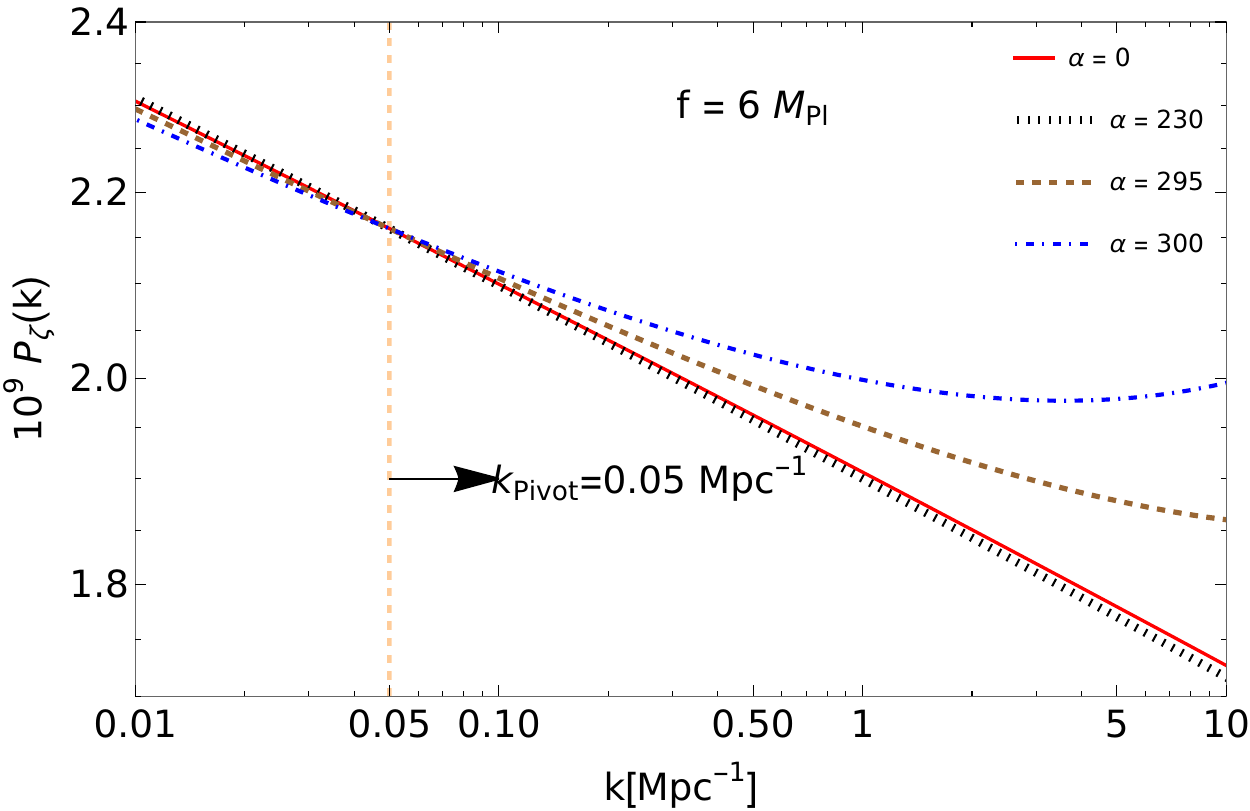}
\vskip -0.1in
\caption{Amplitude of scalar power spectrum around the pivot scale for different values of the coupling constant, $\alpha$ for $f = 6$.}
\label{fig:scalar_power_sp_around_pivot}
\end{figure}

The turn of the spectral index happens due to a competition between the vacuum contribution and the gauge field contribution to the scalar spectrum. For the vacuum part of the scalar power spectrum, the slope decreases at the pivot scale as the coupling constant $\alpha$ increases. This decrease in slope occurs because the pivot scale shifts, which is essentially caused by an increase in the duration of inflation. However, the slope of the scalar power spectrum due to the contributions of the gauge field increases with the coupling constant $\alpha$. At larger values of $\alpha$ the spectral index even becomes greater than $1$ where the gauge fields mainly source the scalar power spectrum \cite{Anber:2009ua}. This can be well understood in Fig.~\ref{fig:scalar_power_sp_around_pivot} where we show the amplitude of the power spectrum only around the pivot scale. We see that the slope makes a turn as we increase $\alpha$. To make this plot, we have adjusted the amplitude at the pivot scale to be the same for all values of $\alpha$ by changing the value $\Lambda$. For the allowed range of $\alpha$, $r$ does not change appreciably and remains within the observational limit for the combined data. 

\begin{figure}[t]
\hspace{-0.5cm}
\includegraphics[width=3.7in]{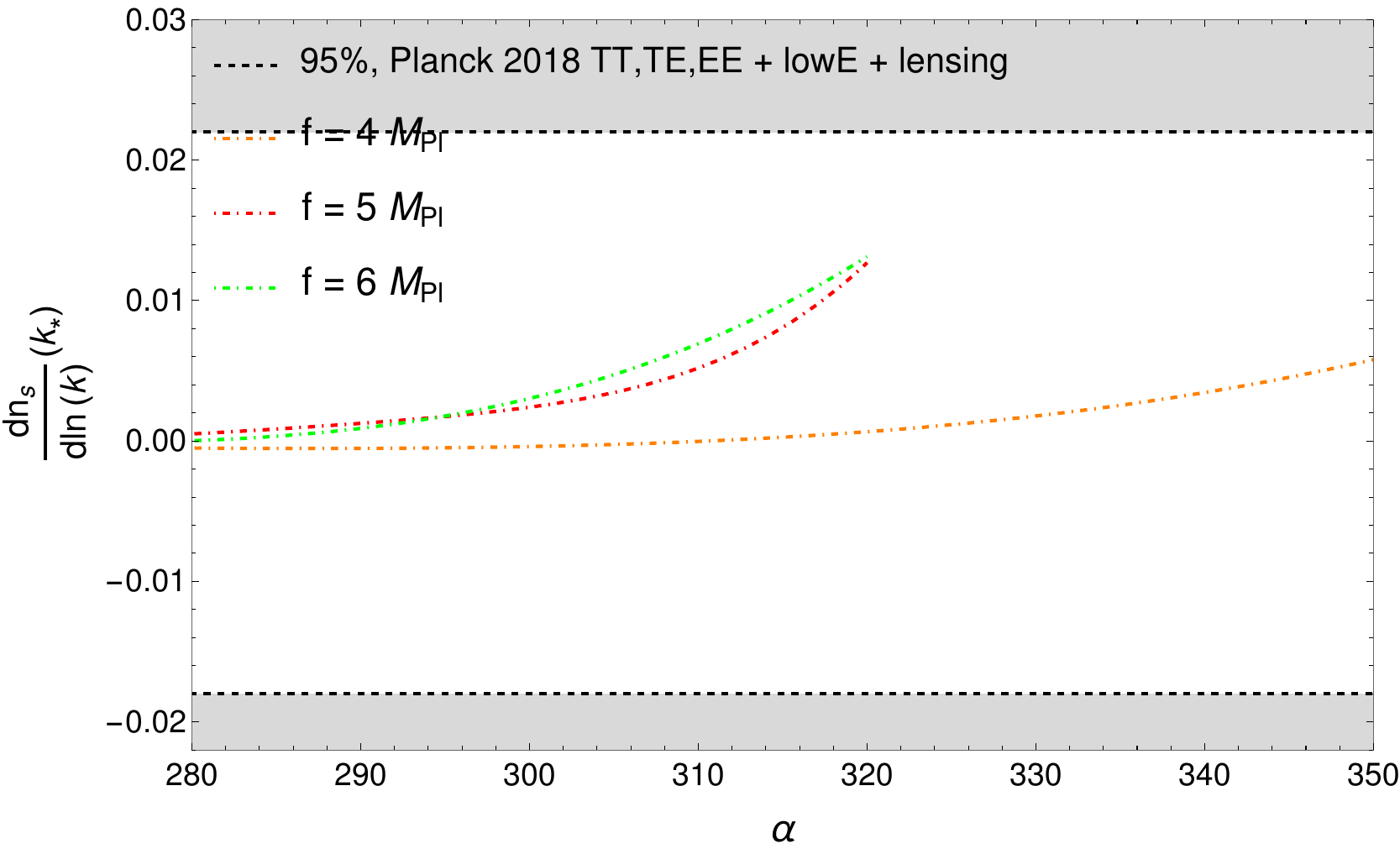}
\includegraphics[width=3.45in]{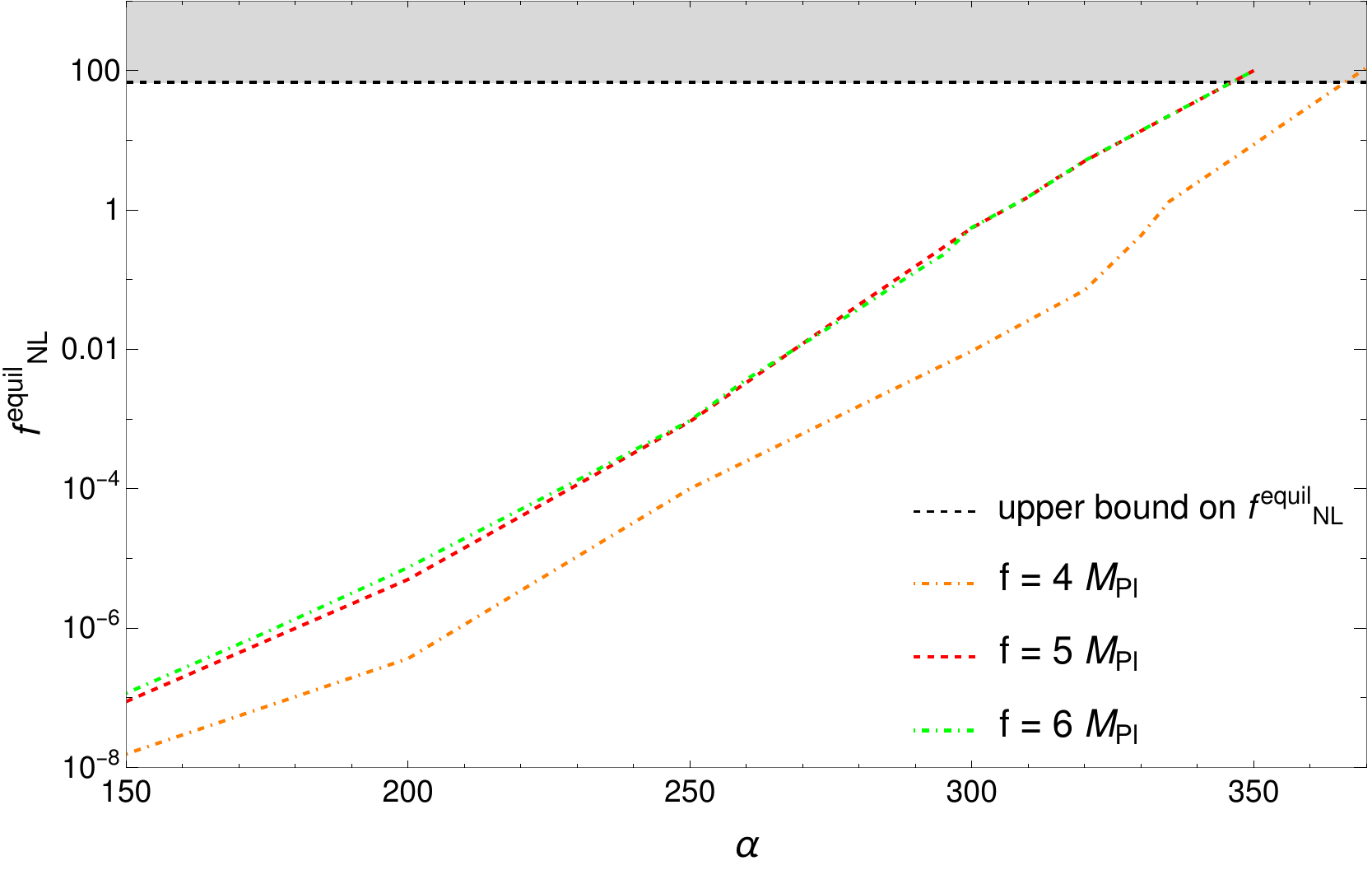}
\vskip -0.1in
\caption{The left panel shows the running of scalar spectral index vs the coupling constant $\alpha$ for $f = 4,~5~\text{and}~ 6$. The shaded regions are excluded by Planck18 data \cite{Akrami:2018odb}. The right panel illustrates the non-causality parameter vs $\alpha$ for $f = 4,~5~\text{and}~ 6$. The horizontal line shows the upper limit from observations \cite{Planck:2019kim}.}
\label{fig:running_of_the_spectral_index_and_non_gausanity}
\end{figure}

As mentioned before, the gauge field not only affect the $n_s$ and $r$, it also induce non-gaussianities and running in the spectrum. Therefore, it is important to check whether the above-mentioned allowed range for $\alpha$ is consistent with the observational limits on running of the spectral index and the non-gaussianities. In the left panel of  Fig.~\ref{fig:running_of_the_spectral_index_and_non_gausanity}, we show the running of the spectral index for different values of the coupling constant $\alpha$ for $f = 4,~ 5$ and $6$. We observe that the running of the spectral index grows with $\alpha$ due to the growing contributions from the gauge field. We also show the $2\sigma$ limit on the running $dn_s/d\ln k= 0.002\pm 0.020$ coming from the Planck18 data \cite{Akrami:2018odb}.
In particular, we find that the allowed range of $\alpha$ in the $n_{s}$-$r$ plane is also within the observational bound provided by the running of the spectral index. Therefore, we find that the values of $\alpha$ for which the model is consistent with $n_s$-$r$ observations are also consistent with running. Note that we have plotted the running of the spectrum up to certain values of $\alpha$ as larger values of $\alpha$ are excluded only by the $n_s$ constraints. 

We now turn our attention to the constraints coming from the non-gassuanities. In this model, the non-gaussianity is of the equilateral type, and it can be calculated by using the formula \cite{Barnaby:2011vw, Barnaby:2010vf}
\begin{equation}
    f^{\rm equil}_{\rm NL}=\left.\frac{f_{3}(\xi)\left(P_{\zeta}(k)_{\rm vac}\right)^3\,e^{6\,\pi\,\xi}}{\left(P_{\zeta}(k)\right)^2}\right\vert_{k=k_*} \label{fnl_formula},
\end{equation}
where $P_{\zeta}(k)_{\rm vac}$ and $P_{\zeta}(k)$ are given in the Eq.~\eqref{scalar_power_sp}, and 
\begin{equation}
    f_{3}(\xi)= \frac{2.8\times 10^{-7}}{\xi^9}~,
\end{equation}
for $\xi\gg 1$.
Now, we plug the numerically evaluated values of $P_{\zeta}(k)_{\rm vac}$, $P_{\zeta}(k)$ and $\xi$ at the pivot scale to get the value of $f^{\rm equil}_{\rm NL}$. It is shown in the right panel of Fig.~\ref{fig:running_of_the_spectral_index_and_non_gausanity} for three values of $f$. The orange dotted line shows the predicted non-gaussianities for different values of $\alpha$ for $f = 4$. The horizontal line is the observational upper limit $ f_{\rm NL}^{\rm equil }=-26\pm 94$ \cite{Planck:2019kim}. We see that all relevant values of $\alpha$ that are consistent with $n_s$ are also consistent with the non-gaussianity bound. We can summarise the result for $f = 4$ by noting that the values $\alpha$ allowed by $n_s$-$r$ constraint, i.e. $347 \lesssim \alpha \lesssim 352$ ($2$-$\sigma$ Planck18+BK18), are also consistent with running and non-gausianities. The constraint on $\alpha$ can be translated to the value $\xi$ at the CMB scale, and for $f = 4$, it is $2.38\lesssim \xi_{*}\lesssim 2.44$.

The constraints on $\alpha$ and therefore $\xi_{*}$ are summarised in Table.~\ref{table_results_N_60_f_1.2} for several values of $f$ and $N_*$ when only Planck18 data is considered. For the case where $f = 5$ and $N_* = 60$, we find that the model is consistent only with Planck18 data at the 2-$\sigma$ level for the parameter values in the range $300 \lesssim \alpha \lesssim 313$. The corresponding values for $\xi_{*}$ fall within the range $2.32 \lesssim \xi_{*} \lesssim 2.44$. We see that for $f = 6$, there is only an upper limit on $\alpha = 305$. In fact, the existence of an upper limit continues for larger values of $f$ also as long as the corresponding $r$ does not become too large. On the other hand for $f\lesssim 5.5$, the model is consistent only for a range of $\alpha$ or $\xi_{*}$. If $f$ becomes smaller than the Planck mass, it is difficult to achieve successful inflation of the required duration of inflation. When we consider $N_*= 50$ (f = 5), the required value of $\alpha$ to make the model consistent with the data is smaller. This happens mainly due to the improved limit on the tensor-to-scalar ratio when two data sets are combined.

Let us put the above result in the context of the previous analysis. In reference, \cite{Barnaby:2011qe}, $\xi_{*}<2.5$ has been derived from the upper bound on the non-gausanity. In the work of \cite{Meerburg:2012id}, the authors did MCMC analysis with the WMAP and ACT data to find an upper limit of $\xi_{*} \lesssim 2.2$ ($95 \%$ CL) for log-flat prior. The constraint came from the violation of scale invariance of the spectrum of Eq.~\eqref{scalar_amplitude} and the constraint on the bispectrum from the WMAP data - see Eq.~\eqref{fnl_formula}. Using the WMAP data, the upper bound on $\xi_{*}$ is also discussed in detail  \cite{Barnaby:2010vf, Barnaby:2011vw, Sorbo:2011rz}.
\begin{table}[!htb]
\centering
\begin{tabular}
{|p{2cm}|p{3.cm}|p{3.4cm}|p{3.5cm}|p{3.4cm}|}
\hline
 & $f=6, N_{*}=60$ &$f=5, N_{*} = 60$ &$f=4, N_{*}=60$&$f=5, N_{*}=50$\\
\hline
$\alpha$ values & ~~~~~$\alpha\leq 305$ &~~~~~$300\leq\alpha\leq313$   & ~~~~$342\leq\alpha\leq352$ &~~~~$220\leq\alpha\leq230$\\
\hline
$\xi_{*}$ values & $~~~~\xi_{*}< 2.25$ & ~~2.23$\leq\xi_{*}\leq$2.36  & ~~$2.34\leq\xi_{*}\leq 2.44$ & $2.26\leq\xi_{*}\leq 2.34$\\
\hline
\end{tabular}
\caption{Constraints on $\alpha$ and $\xi_{*}$ for several values of $f$ and $N_*$ when only Planck18 data is used. }
\label{table_results_N_60_f_1.2}
\end{table}

In summary, pNGB inflation with periodic potential is under severe tensions when it is confronted with the data. We find out that the coupling between the pNGB and a gauge field can ameliorate the tension only for superPlanckian values of $f$. In this case, we find that the recent constraints coming from $n_s$ and $r$ are more severe than the non-gaussinaities and running of the spectral index as were found earlier. Interestingly, this leads to a range of consistent $\alpha$ or $\xi_{*}$ in contrast to the previous analysis where only an upper limit existed.

\section{Discussions and conclusion }
\label{Sec:Conclusion}

In this work, we have investigated that the gauge field is produced due to Chern-Simons-like parity-violating coupling between the pseudoscalar inflaton field and a gauge field. We have focused our attention on studying the back-reaction effects on the background dynamics. Confirming previous works, our analysis demonstrates that the back-reaction significantly influences the inflation dynamics, leading to notable changes in the duration of the inflationary period and the estimation of the number of $e$-foldings $N$. This prolonged inflation duration affects the predictions of the CMB spectral indices at the pivot scale. The main findings of our paper can be summarized as follows.

The duration of inflation is prolonged due to the production of gauge fields via Chern-Simons-like coupling, as depicted in Fig.~\ref{fig:epsilon1}. If we consider that the pivot mode exits the horizon at a fixed value of $e$-folding ($N_{*}=60$ or $N_{*}=50$) before the end of inflation, then due to the varied amounts of the back-reactions (for different values of $\alpha$), the CMB modes exits the horizon while the field is in different parts of the inflationary potential. In this way, the coupling constant $\alpha$ affects the CMB observables (see Fig.~\ref{fig:inflaton_dynmics})). From Eq.~\eqref{eq:pertScl2}, we observe that the production of the gauge field also acts as a source term for the inhomogeneous dynamics of the inflaton. Due to the growth of this source term during inflation, we observe a growth in the scalar power spectrum at small scales, as illustrated in the left panel of Fig. \ref{fig:scalar_tensor_power_spectra}. As $\alpha$ increases, the effects of the gauge field on the scalar power spectra start to affect the large scale also (refer to the left panel of Fig. \ref{fig:scalar_tensor_power_spectra}). The scalar power spectrum exhibits oscillatory features at small scales because of the strong back-reaction effects in the later stages of inflation, and it does not diverge because we have used the corrected power spectrum expression of Eq.~\eqref{scalar_power_sp}, which includes the effect of strong back-reactions. Similarly, we computed the tensor power spectrum and noted an increase in its magnitude at the small scale due to the presence of the gauge field.

The predictions of the CMB spectral index $n_{s}$ and the tensor-to-scalar ratio $r$ for the axion model are under tension with current observational data - see the continuous line in the Fig.~\ref{fig:N_s-vs-r.pdf}. When both Planck18 and BK18 data are used, the model is ruled out at $2$-$\sigma$, whereas it is consistent at $2$-$\sigma$ for $f \gtrsim 5.5$ when only Planck18 data is considered. However, the presence of the gauge field affects the CMB scales, leading to significant impacts on the predictions of the spectral index ($n_s$) and the tensor-to-scalar ratio (r) at the CMB scales. As we increase the value of the coupling constant $\alpha$, the spectral index starts to decrease first, but with further increases of the $\alpha$, the spectral index starts to reverse its trend. This seems to be the common feature for all values of $f$. Therefore, for all super-Planckian values of $f$, the model becomes consistent (in reference to different data combinations) for a certain range of $\alpha$ only. Natural inflation model with no gauge field is excluded at $2$-$\sigma$ for all values of $f$ when viewed w.r.t Planck18+BK18 data. But, for $f = 4$, the model becomes consistent for a range of values of $\alpha$ for Planck18+BK18 data. Interestingly enough, the constraints coming from $n_s$-$r$ plane are more stringent than the constraints originating from the non-gaussianities or running of the spectrum. In comparison to the previous analysis where an upper limit was derived, we find a range of $\xi_*$, or $\alpha$ for which the model is consistent. The existence of both lower and upper limits originates due to the improved limit of $n_s$ from the latest data. The limits are summarised in Table~ \ref{table_results_N_60_f_1.2}. In short, a coupling between the gauge field and the pseudoscalar inflaton helps to alleviate the tension of the natural inflation model and makes it consistent with the current observational data.

 As a future outlook,  it is quite natural to extend our analysis to non-Abelian and massive gauge fields, where self-interactions among the gauge fields may have several phenomenological implications. It may lead to a wider range of consistent values of the coupling constant $\alpha$. The crucial parameter is the mass of the gauge field relative to the Hubble parameter. 
 In our present work, we solved the background dynamics of the inflaton field numerically, where the gauge field acts as a source term. However, we have not numerically solved the inhomogeneous differential equation of the inflaton field in the presence of the gauge field or the scalar power spectrum of the inflaton field. To solve the inhomogeneous equation, we have to use lattice code as outlined here \cite{Figueroa:2023oxc}. More precise constraints on the parameters must consider these effects.

 In our analysis, we have parameterized the post-inflationary uncertainty by using two values of $N_*$. In principle, $N_*$ should be taken as a variable in estimating the best fit parameter for the model. The analysis can be carried out along the lines of work \cite{Bhattacharya:2017pws, Bhattacharya:2020gnk, Bhattacharya:2022akq} where other than the usual parameters, three model parameters, namely $\Lambda, f, \alpha$, and $N_*$ should be used in the MCMC analysis. Another interesting avenue is to explore the production of primordial black holes and secondary gravitational waves due to the enhanced power spectrum at small scales. Our results show that the model can be consistent only for large values of $f$ with a certain range for $\alpha$. Therefore, the associated predictions for primordial black holes and gravitational waves will also be in specific mass ranges and frequencies respectively. We wish to explore some of the aspects of the model in our future work.

In summary, we have explored the canonical natural inflation model with a periodic cosine potential in the presence of a gauge field. Due to the coupling between the gauge field and the inflaton, gauge quanta are excited leading to large scalar fluctuations at the later stage of inflation. We have reexamined the set-up in light of the latest CMB data. Our finding shows that for super-Planckian $f$ values, the model can be consistent with observations for a narrow range of the coupling parameter $\alpha$, whereas without any coupling the model is under severe tension.

\medskip
\section*{Acknowledgement}
The authors would like to thank Anish Ghoshal for several fruitful discussions, and for taking part in the initial stages of the project. N.J thanks Shu-Lin Cheng for his help in numerical simulations. N.J. was supported by the National Postdoctoral Fellowship of the Science and Engineering Research Board (SERB),
Department of Science and Technology (DST), Government of India, File No. PDF/2021/004114. K.D would like to acknowledge support from the ICTP, Trieste through the Associates Programme (2023-2028). We thank the anonymous referee for her/his critical comments about the importance of the results, and accordingly, we have presented our results slightly differently. 

\appendix

\section{Numerical methods}
\label{appendix1}
In this appendix, we discuss in detail how we numerically solve the background dynamics of the inflaton field and the Hubble parameter, taking into account the dynamical back-reaction term \cite{Cheng:2015oqa}. In our case, we consider the effects of the gauge field both in the equation of motion of the inflation field and in the Hubble equation. We have to solve simultaneously the time evolution of the gauge field mode of Eq.~\eqref{mode_of_gauge_fld}, inflaton dynamics of  Eq.~\eqref{backr1}, and the Hubble equation~\eqref{backr2} using the expressions in Eq.~\eqref{edb_equation} and Eq.~\eqref{energy_of_EM_fld}. For this purpose, we employ the fourth-order Runge-Kutta method, commonly denoted as \textit{RK4}. In the following, we outline the methodology. 

First, for $\alpha = 0$ and a chosen value of $f$, we solve the background equation for the inflaton field and the scale factor. With the initial field velocity set at the slow-roll velocity, we choose the initial field value such that the required number of e-foldings of inflation is achieved. The value of $\Lambda$ is fixed by the amplitude of scalar perturbations at the CMB scales. When we do the same exercise for $\alpha \neq 0$, the duration of inflation is elongated and the amplitude is calculated at different points of the inflaton potential, and accordingly $\Lambda$ is readjusted such that its value remains within the observed limit of the scalar amplitude. 

At the initial time $t_i$, the gauge field has only vacuum fluctuations, contributing nothing to the back-reaction terms. For the next time interval ($t_i+h$), we evaluate  $\phi_0(t_i+h)$ and $\dot{\phi_0}(t_{i}+h)$ by solving Eq.~\eqref{backr1} and Eq.~\eqref{backr2}. In addition, we simultaneosuly solve Eq.~\eqref{mode_of_gauge_fld} for all values of $k$ to find out the expression for Eq.~\eqref{energy_of_EM_fld} and Eq.~\eqref{edb_equation}. To solve, it is assumed that the gauges field modes are in the Bunch Davies vacua with the following initial values:
\begin{eqnarray}
    A^{\pm}_{\rm real} &= & \frac{1}{\sqrt{2k}} \, \, ,   A^{\pm}_{\rm im}=  0~,\\ 
    \frac{d A^{\pm}_{\rm real}}{dt} &= & 0\, ,\, \frac{dA^{\pm}_{\rm im  }}{dt} =  -\sqrt{\frac{k}{2}} \frac{1}{a} \, \,  . 
\end{eqnarray}
At the time ($t_i+2h$), the evaluation of the inflaton field is calculated with the source term. By continuing the process, we find all relevant dynamical quantities over a period of time that ends with the $\epsilon_H = 1$. 

Now, we will discuss how to evaluate the source term in the inflation equation of motion and the additional friction term in the Hubble equation by solving  Eq.~\eqref{mode_of_gauge_fld}. With our convention of $\dot \phi_{0}  > 0$ at the initial stages of inflation, it remains positive throughout its evolution even during the strong back-reaction regime at the last stage of inflation. Therefore $\xi$ remains always positive. In this case, $A^-$ mode always oscillates for all values of $k$. On the other hand, $A^+$ mode grows exponentially for $k < 2 aH \xi $, and oscillates for $k > 2 aH \xi$.
For a value $\xi$ at any moment, the $A^+$ modes that satisfy the above-mentioned instability conditions are numerically evolved to the next step, but other modes are reset to their corresponding Bunch Davies vacuum. We keep $A^-_k$ modes at their vacuum values. It allows us to evaluate the terms of Eq.~\eqref{energy_of_EM_fld} and Eq.~\eqref{edb_equation} at that moment after doing the momentum integration. In the next of the numerical evolution, the inflaton field and the Hubble equation receive the back-reaction effects due to the gauge field. As $\xi$ increases, more number of modes get excited and contribute to the integration. As seen in Fig.~\ref{fig:xi}, at the later stages of inflation, $\xi$ shows oscialltory behaviour. Therefore, when $\xi$ values decreases from its local maximum values, lesser number of modes get exponentially excited. At that point of time, the modes which were excited earlier but not now, are kept at the previous values.  
 
\bibliographystyle{JHEP}
\bibliography{References.bib}

\providecommand{\href}[2]{#2}\begingroup\raggedright\begin{thebibliography}{10}

\bibitem{Akrami:2018odb}
{\bf Planck} Collaboration, Y.~Akrami et~al., {\it {Planck 2018 results. X.
  Constraints on inflation}},  {\em Astron. Astrophys.} {\bf 641} (2020) A10,
  [\href{http://arxiv.org/abs/1807.06211}{{\tt arXiv:1807.06211}}].

\bibitem{Guth:1980zm}
A.~H. Guth, {\it {The Inflationary Universe: A Possible Solution to the Horizon
  and Flatness Problems}},  {\em Phys. Rev. D} {\bf 23} (1981) 347--356.

\bibitem{Starobinsky:1980te}
A.~A. Starobinsky, {\it {A New Type of Isotropic Cosmological Models Without
  Singularity}},  {\em Phys. Lett. B} {\bf 91} (1980) 99--102.

\bibitem{Linde:1981mu}
A.~D. Linde, {\it {A New Inflationary Universe Scenario: A Possible Solution of
  the Horizon, Flatness, Homogeneity, Isotropy and Primordial Monopole
  Problems}},  {\em Phys. Lett. B} {\bf 108} (1982) 389--393.

\bibitem{Albrecht:1982wi}
A.~Albrecht and P.~J. Steinhardt, {\it {Cosmology for Grand Unified Theories
  with Radiatively Induced Symmetry Breaking}},  {\em Phys. Rev. Lett.} {\bf
  48} (1982) 1220--1223.

\bibitem{Martin:2013tda}
J.~Martin, C.~Ringeval, and V.~Vennin, {\it {Encyclop\ae{}dia Inflationaris}},
  {\em Phys. Dark Univ.} {\bf 5-6} (2014) 75--235,
  [\href{http://arxiv.org/abs/1303.3787}{{\tt arXiv:1303.3787}}].

\bibitem{Dutta:2021bih}
K.~Dutta and A.~Maharana, {\it {Models of accelerating universe in supergravity
  and string theory}},  {\em Eur. Phys. J. ST} {\bf 230} (2021), no.~9
  2111--2122.

\bibitem{Baumann:2014nda}
D.~Baumann and L.~McAllister, {\em {Inflation and String Theory}}.
\newblock Cambridge Monographs on Mathematical Physics. Cambridge University
  Press, 5, 2015.

\bibitem{McAllister:2008hb}
L.~McAllister, E.~Silverstein, and A.~Westphal, {\it {Gravity Waves and Linear
  Inflation from Axion Monodromy}},  {\em Phys. Rev. D} {\bf 82} (2010) 046003,
  [\href{http://arxiv.org/abs/0808.0706}{{\tt arXiv:0808.0706}}].

\bibitem{Kaloper:2008fb}
N.~Kaloper and L.~Sorbo, {\it {A Natural Framework for Chaotic Inflation}},
  {\em Phys. Rev. Lett.} {\bf 102} (2009) 121301,
  [\href{http://arxiv.org/abs/0811.1989}{{\tt arXiv:0811.1989}}].

\bibitem{Freese:1990rb}
K.~Freese, J.~A. Frieman, and A.~V. Olinto, {\it {Natural inflation with pseudo
  - Nambu-Goldstone bosons}},  {\em Phys. Rev. Lett.} {\bf 65} (1990)
  3233--3236.

\bibitem{Adams:1992bn}
F.~C. Adams, J.~R. Bond, K.~Freese, J.~A. Frieman, and A.~V. Olinto, {\it
  {Natural inflation: Particle physics models, power law spectra for large
  scale structure, and constraints from COBE}},  {\em Phys. Rev. D} {\bf 47}
  (1993) 426--455.

\bibitem{BICEP:2021xfz}
{\bf BICEP, Keck} Collaboration, P.~A.~R. Ade et~al., {\it {Improved
  Constraints on Primordial Gravitational Waves using Planck, WMAP, and
  BICEP/Keck Observations through the 2018 Observing Season}},  {\em Phys. Rev.
  Lett.} {\bf 127} (2021), no.~15 151301,
  [\href{http://arxiv.org/abs/2110.00483}{{\tt arXiv:2110.00483}}].

\bibitem{Arkani-Hamed:2003xts}
N.~Arkani-Hamed, H.-C. Cheng, P.~Creminelli, and L.~Randall, {\it {Extra
  natural inflation}},  {\em Phys. Rev. Lett.} {\bf 90} (2003) 221302,
  [\href{http://arxiv.org/abs/hep-th/0301218}{{\tt hep-th/0301218}}].

\bibitem{Kallosh:1995hi}
R.~Kallosh, A.~D. Linde, D.~A. Linde, and L.~Susskind, {\it {Gravity and global
  symmetries}},  {\em Phys. Rev. D} {\bf 52} (1995) 912--935,
  [\href{http://arxiv.org/abs/hep-th/9502069}{{\tt hep-th/9502069}}].

\bibitem{Salvio:2023cry}
A.~Salvio and S.~Sciusco, {\it {(Multi-field) natural inflation and
  gravitational waves}},  {\em JCAP} {\bf 03} (2024) 018,
  [\href{http://arxiv.org/abs/2311.00741}{{\tt arXiv:2311.00741}}].

\bibitem{Banks:2003sx}
T.~Banks, M.~Dine, P.~J. Fox, and E.~Gorbatov, {\it {On the possibility of
  large axion decay constants}},  {\em JCAP} {\bf 06} (2003) 001,
  [\href{http://arxiv.org/abs/hep-th/0303252}{{\tt hep-th/0303252}}].

\bibitem{Kim:2004rp}
J.~E. Kim, H.~P. Nilles, and M.~Peloso, {\it {Completing natural inflation}},
  {\em JCAP} {\bf 01} (2005) 005,
  [\href{http://arxiv.org/abs/hep-ph/0409138}{{\tt hep-ph/0409138}}].

\bibitem{Dimopoulos:2005ac}
S.~Dimopoulos, S.~Kachru, J.~McGreevy, and J.~G. Wacker, {\it {N-flation}},
  {\em JCAP} {\bf 08} (2008) 003,
  [\href{http://arxiv.org/abs/hep-th/0507205}{{\tt hep-th/0507205}}].

\bibitem{Cicoli:2014sva}
M.~Cicoli, K.~Dutta, and A.~Maharana, {\it {N-flation with Hierarchically Light
  Axions in String Compactifications}},  {\em JCAP} {\bf 08} (2014) 012,
  [\href{http://arxiv.org/abs/1401.2579}{{\tt arXiv:1401.2579}}].

\bibitem{Das:2014gua}
K.~Das and K.~Dutta, {\it {N-flation in Supergravity}},  {\em Phys. Lett. B}
  {\bf 738} (2014) 457--463, [\href{http://arxiv.org/abs/1408.6376}{{\tt
  arXiv:1408.6376}}].

\bibitem{Garretson:1992vt}
W.~D. Garretson, G.~B. Field, and S.~M. Carroll, {\it {Primordial magnetic
  fields from pseudoGoldstone bosons}},  {\em Phys. Rev. D} {\bf 46} (1992)
  5346--5351, [\href{http://arxiv.org/abs/hep-ph/9209238}{{\tt
  hep-ph/9209238}}].

\bibitem{Anber:2006xt}
M.~M. Anber and L.~Sorbo, {\it {N-flationary magnetic fields}},  {\em JCAP}
  {\bf 10} (2006) 018, [\href{http://arxiv.org/abs/astro-ph/0606534}{{\tt
  astro-ph/0606534}}].

\bibitem{Ballardini:2019rqh}
M.~Ballardini, M.~Braglia, F.~Finelli, G.~Marozzi, and A.~A. Starobinsky, {\it
  {Energy-momentum tensor and helicity for gauge fields coupled to a
  pseudo-scalar inflaton}},  {\em Phys. Rev. D} {\bf 100} (2019), no.~12
  123542, [\href{http://arxiv.org/abs/1910.13448}{{\tt arXiv:1910.13448}}].
  [Erratum: Phys.Rev.D 105, 069905 (2022)].

\bibitem{Campeti:2022acx}
P.~Campeti, O.~\"Ozsoy, I.~Obata, and M.~Shiraishi, {\it {New constraints on
  axion-gauge field dynamics during inflation from Planck and BICEP/Keck data
  sets}},  {\em JCAP} {\bf 07} (2022), no.~07 039,
  [\href{http://arxiv.org/abs/2203.03401}{{\tt arXiv:2203.03401}}].

\bibitem{Anber:2009ua}
M.~M. Anber and L.~Sorbo, {\it {Naturally inflating on steep potentials through
  electromagnetic dissipation}},  {\em Phys. Rev. D} {\bf 81} (2010) 043534,
  [\href{http://arxiv.org/abs/0908.4089}{{\tt arXiv:0908.4089}}].

\bibitem{Barnaby:2011qe}
N.~Barnaby, E.~Pajer, and M.~Peloso, {\it {Gauge Field Production in Axion
  Inflation: Consequences for Monodromy, non-Gaussianity in the CMB, and
  Gravitational Waves at Interferometers}},  {\em Phys. Rev. D} {\bf 85} (2012)
  023525, [\href{http://arxiv.org/abs/1110.3327}{{\tt arXiv:1110.3327}}].

\bibitem{Barnaby:2010vf}
N.~Barnaby and M.~Peloso, {\it {Large Nongaussianity in Axion Inflation}},
  {\em Phys. Rev. Lett.} {\bf 106} (2011) 181301,
  [\href{http://arxiv.org/abs/1011.1500}{{\tt arXiv:1011.1500}}].

\bibitem{Barnaby:2011vw}
N.~Barnaby, R.~Namba, and M.~Peloso, {\it {Phenomenology of a Pseudo-Scalar
  Inflaton: Naturally Large Nongaussianity}},  {\em JCAP} {\bf 04} (2011) 009,
  [\href{http://arxiv.org/abs/1102.4333}{{\tt arXiv:1102.4333}}].

\bibitem{Linde:2012bt}
A.~Linde, S.~Mooij, and E.~Pajer, {\it {Gauge field production in supergravity
  inflation: Local non-Gaussianity and primordial black holes}},  {\em Phys.
  Rev. D} {\bf 87} (2013), no.~10 103506,
  [\href{http://arxiv.org/abs/1212.1693}{{\tt arXiv:1212.1693}}].

\bibitem{Bugaev:2013fya}
E.~Bugaev and P.~Klimai, {\it {Axion inflation with gauge field production and
  primordial black holes}},  {\em Phys. Rev. D} {\bf 90} (2014), no.~10 103501,
  [\href{http://arxiv.org/abs/1312.7435}{{\tt arXiv:1312.7435}}].

\bibitem{Cheng:2015oqa}
S.-L. Cheng, W.~Lee, and K.-W. Ng, {\it {Numerical study of pseudoscalar
  inflation with an axion-gauge field coupling}},  {\em Phys. Rev. D} {\bf 93}
  (2016), no.~6 063510, [\href{http://arxiv.org/abs/1508.00251}{{\tt
  arXiv:1508.00251}}].

\bibitem{Garcia-Bellido:2016dkw}
J.~Garcia-Bellido, M.~Peloso, and C.~Unal, {\it {Gravitational waves at
  interferometer scales and primordial black holes in axion inflation}},  {\em
  JCAP} {\bf 12} (2016) 031, [\href{http://arxiv.org/abs/1610.03763}{{\tt
  arXiv:1610.03763}}].

\bibitem{Domcke:2017fix}
V.~Domcke, F.~Muia, M.~Pieroni, and L.~T. Witkowski, {\it {PBH dark matter from
  axion inflation}},  {\em JCAP} {\bf 07} (2017) 048,
  [\href{http://arxiv.org/abs/1704.03464}{{\tt arXiv:1704.03464}}].

\bibitem{Garcia-Bellido:2017aan}
J.~Garcia-Bellido, M.~Peloso, and C.~Unal, {\it {Gravitational Wave signatures
  of inflationary models from Primordial Black Hole Dark Matter}},  {\em JCAP}
  {\bf 09} (2017) 013, [\href{http://arxiv.org/abs/1707.02441}{{\tt
  arXiv:1707.02441}}].

\bibitem{Cheng:2018yyr}
S.-L. Cheng, W.~Lee, and K.-W. Ng, {\it {Primordial black holes and associated
  gravitational waves in axion monodromy inflation}},  {\em JCAP} {\bf 07}
  (2018) 001, [\href{http://arxiv.org/abs/1801.09050}{{\tt arXiv:1801.09050}}].

\bibitem{Ozsoy:2023ryl}
O.~\"Ozsoy and G.~Tasinato, {\it {Inflation and Primordial Black Holes}},  {\em
  Universe} {\bf 9} (2023), no.~5 203,
  [\href{http://arxiv.org/abs/2301.03600}{{\tt arXiv:2301.03600}}].

\bibitem{Cook:2022zol}
J.~L. Cook, {\it {Primordial black hole production in natural and hilltop
  inflation}},  {\em JCAP} {\bf 07} (2023) 031,
  [\href{http://arxiv.org/abs/2209.05674}{{\tt arXiv:2209.05674}}].

\bibitem{Sorbo:2011rz}
L.~Sorbo, {\it {Parity violation in the Cosmic Microwave Background from a
  pseudoscalar inflaton}},  {\em JCAP} {\bf 06} (2011) 003,
  [\href{http://arxiv.org/abs/1101.1525}{{\tt arXiv:1101.1525}}].

\bibitem{Cook:2011hg}
J.~L. Cook and L.~Sorbo, {\it {Particle production during inflation and
  gravitational waves detectable by ground-based interferometers}},  {\em Phys.
  Rev. D} {\bf 85} (2012) 023534, [\href{http://arxiv.org/abs/1109.0022}{{\tt
  arXiv:1109.0022}}]. [Erratum: Phys.Rev.D 86, 069901 (2012)].

\bibitem{Anber:2012du}
M.~M. Anber and L.~Sorbo, {\it {Non-Gaussianities and chiral gravitational
  waves in natural steep inflation}},  {\em Phys. Rev. D} {\bf 85} (2012)
  123537, [\href{http://arxiv.org/abs/1203.5849}{{\tt arXiv:1203.5849}}].

\bibitem{Domcke:2016bkh}
V.~Domcke, M.~Pieroni, and P.~Bin\'etruy, {\it {Primordial gravitational waves
  for universality classes of pseudoscalar inflation}},  {\em JCAP} {\bf 06}
  (2016) 031, [\href{http://arxiv.org/abs/1603.01287}{{\tt arXiv:1603.01287}}].

\bibitem{Garcia-Bellido:2023ser}
J.~Garcia-Bellido, A.~Papageorgiou, M.~Peloso, and L.~Sorbo, {\it {A flashing
  beacon in axion inflation: recurring bursts of gravitational waves in the
  strong backreaction regime}},  {\em JCAP} {\bf 01} (2024) 034,
  [\href{http://arxiv.org/abs/2303.13425}{{\tt arXiv:2303.13425}}].

\bibitem{LISACosmologyWorkingGroup:2023njw}
{\bf LISA Cosmology Working Group} Collaboration, E.~Bagui et~al., {\it
  {Primordial black holes and their gravitational-wave signatures}},
  \href{http://arxiv.org/abs/2310.19857}{{\tt arXiv:2310.19857}}.

\bibitem{Caprini:2014mja}
C.~Caprini and L.~Sorbo, {\it {Adding helicity to inflationary
  magnetogenesis}},  {\em JCAP} {\bf 10} (2014) 056,
  [\href{http://arxiv.org/abs/1407.2809}{{\tt arXiv:1407.2809}}].

\bibitem{Adshead:2016iae}
P.~Adshead, J.~T. Giblin, T.~R. Scully, and E.~I. Sfakianakis, {\it
  {Magnetogenesis from axion inflation}},  {\em JCAP} {\bf 10} (2016) 039,
  [\href{http://arxiv.org/abs/1606.08474}{{\tt arXiv:1606.08474}}].

\bibitem{Jimenez:2017cdr}
D.~Jim\'enez, K.~Kamada, K.~Schmitz, and X.-J. Xu, {\it {Baryon asymmetry and
  gravitational waves from pseudoscalar inflation}},  {\em JCAP} {\bf 12}
  (2017) 011, [\href{http://arxiv.org/abs/1707.07943}{{\tt arXiv:1707.07943}}].

\bibitem{Domcke:2019mnd}
V.~Domcke, B.~von Harling, E.~Morgante, and K.~Mukaida, {\it {Baryogenesis from
  axion inflation}},  {\em JCAP} {\bf 10} (2019) 032,
  [\href{http://arxiv.org/abs/1905.13318}{{\tt arXiv:1905.13318}}].

\bibitem{Domcke:2018eki}
V.~Domcke and K.~Mukaida, {\it {Gauge Field and Fermion Production during Axion
  Inflation}},  {\em JCAP} {\bf 11} (2018) 020,
  [\href{http://arxiv.org/abs/1806.08769}{{\tt arXiv:1806.08769}}].

\bibitem{Talebian:2022cwk}
A.~Talebian, S.~A. Hosseini~Mansoori, and H.~Firouzjahi, {\it {Inflation from
  Multiple Pseudo-scalar Fields: Primordial Black Hole Dark Matter and
  Gravitational Waves}},  {\em Astrophys. J.} {\bf 948} (2023), no.~1 48,
  [\href{http://arxiv.org/abs/2210.13822}{{\tt arXiv:2210.13822}}].

\bibitem{vonEckardstein:2023gwk}
R.~von Eckardstein, M.~Peloso, K.~Schmitz, O.~Sobol, and L.~Sorbo, {\it {Axion
  inflation in the strong-backreaction regime: decay of the Anber-Sorbo
  solution}},  {\em JHEP} {\bf 11} (2023) 183,
  [\href{http://arxiv.org/abs/2309.04254}{{\tt arXiv:2309.04254}}].

\bibitem{Peloso:2022ovc}
M.~Peloso and L.~Sorbo, {\it {Instability in axion inflation with strong
  backreaction from gauge modes}},  {\em JCAP} {\bf 01} (2023) 038,
  [\href{http://arxiv.org/abs/2209.08131}{{\tt arXiv:2209.08131}}].

\bibitem{Notari:2016npn}
A.~Notari and K.~Tywoniuk, {\it {Dissipative Axial Inflation}},  {\em JCAP}
  {\bf 12} (2016) 038, [\href{http://arxiv.org/abs/1608.06223}{{\tt
  arXiv:1608.06223}}].

\bibitem{DallAgata:2019yrr}
G.~Dall'Agata, S.~Gonz\'alez-Mart\'\i{}n, A.~Papageorgiou, and M.~Peloso, {\it
  {Warm dark energy}},  {\em JCAP} {\bf 08} (2020) 032,
  [\href{http://arxiv.org/abs/1912.09950}{{\tt arXiv:1912.09950}}].

\bibitem{Sobol:2019xls}
O.~O. Sobol, E.~V. Gorbar, and S.~I. Vilchinskii, {\it {Backreaction of
  electromagnetic fields and the Schwinger effect in pseudoscalar inflation
  magnetogenesis}},  {\em Phys. Rev. D} {\bf 100} (2019), no.~6 063523,
  [\href{http://arxiv.org/abs/1907.10443}{{\tt arXiv:1907.10443}}].

\bibitem{Cado:2022pxk}
Y.~Cado and M.~Quir\'os, {\it {Numerical study of the Schwinger effect in axion
  inflation}},  {\em Phys. Rev. D} {\bf 106} (2022), no.~12 123527,
  [\href{http://arxiv.org/abs/2208.10977}{{\tt arXiv:2208.10977}}].

\bibitem{Domcke:2020zez}
V.~Domcke, V.~Guidetti, Y.~Welling, and A.~Westphal, {\it {Resonant
  backreaction in axion inflation}},  {\em JCAP} {\bf 09} (2020) 009,
  [\href{http://arxiv.org/abs/2002.02952}{{\tt arXiv:2002.02952}}].

\bibitem{Caravano:2022epk}
A.~Caravano, E.~Komatsu, K.~D. Lozanov, and J.~Weller, {\it {Lattice
  simulations of axion-U(1) inflation}},  {\em Phys. Rev. D} {\bf 108} (2023),
  no.~4 043504, [\href{http://arxiv.org/abs/2204.12874}{{\tt
  arXiv:2204.12874}}].

\bibitem{Figueroa:2023oxc}
D.~G. Figueroa, J.~Lizarraga, A.~Urio, and J.~Urrestilla, {\it {Strong
  Backreaction Regime in Axion Inflation}},  {\em Phys. Rev. Lett.} {\bf 131}
  (2023), no.~15 151003, [\href{http://arxiv.org/abs/2303.17436}{{\tt
  arXiv:2303.17436}}].

\bibitem{Gorbar:2021rlt}
E.~V. Gorbar, K.~Schmitz, O.~O. Sobol, and S.~I. Vilchinskii, {\it {Gauge-field
  production during axion inflation in the gradient expansion formalism}},
  {\em Phys. Rev. D} {\bf 104} (2021), no.~12 123504,
  [\href{http://arxiv.org/abs/2109.01651}{{\tt arXiv:2109.01651}}].

\bibitem{Domcke:2023tnn}
V.~Domcke, Y.~Ema, and S.~Sandner, {\it {Perturbatively including
  inhomogeneities in axion inflation}},  {\em JCAP} {\bf 03} (2024) 019,
  [\href{http://arxiv.org/abs/2310.09186}{{\tt arXiv:2310.09186}}].

\bibitem{Durrer:2024ibi}
R.~Durrer, R.~von Eckardstein, D.~Garg, K.~Schmitz, O.~Sobol, and
  S.~Vilchinskii, {\it {Scalar perturbations from inflation in the presence of
  gauge fields}},  {\em Phys. Rev. D} {\bf 110} (2024), no.~4 043533,
  [\href{http://arxiv.org/abs/2404.19694}{{\tt arXiv:2404.19694}}].

\bibitem{Bird_2011}
S.~Bird, H.~V. Peiris, M.~Viel, and L.~Verde, {\it Minimally parametric power
  spectrum reconstruction from the lyman $\alpha$ forest: P(k) reconstruction
  from lyman $\alpha$},  {\em Monthly Notices of the Royal Astronomical
  Society} {\bf 413} (Mar., 2011) 1717–1728.

\bibitem{A_Kogut_2011}
A.~Kogut, D.~Fixsen, D.~Chuss, J.~Dotson, E.~Dwek, M.~Halpern, G.~Hinshaw,
  S.~Meyer, S.~Moseley, M.~Seiffert, D.~Spergel, and E.~Wollack, {\it The
  primordial inflation explorer (pixie): a nulling polarimeter for cosmic
  microwave background observations},  {\em Journal of Cosmology and
  Astroparticle Physics} {\bf 2011} (July, 2011) 025–025.

\bibitem{Fixsen:1996nj}
D.~J. Fixsen, E.~S. Cheng, J.~M. Gales, J.~C. Mather, R.~A. Shafer, and E.~L.
  Wright, {\it {The Cosmic Microwave Background spectrum from the full COBE
  FIRAS data set}},  {\em Astrophys. J.} {\bf 473} (1996) 576,
  [\href{http://arxiv.org/abs/astro-ph/9605054}{{\tt astro-ph/9605054}}].

\bibitem{Byrnes:2018txb}
C.~T. Byrnes, P.~S. Cole, and S.~P. Patil, {\it {Steepest growth of the power
  spectrum and primordial black holes}},  {\em JCAP} {\bf 06} (2019) 028,
  [\href{http://arxiv.org/abs/1811.11158}{{\tt arXiv:1811.11158}}].

\bibitem{Planck:2019kim}
{\bf Planck} Collaboration, Y.~Akrami et~al., {\it {Planck 2018 results. IX.
  Constraints on primordial non-Gaussianity}},  {\em Astron. Astrophys.} {\bf
  641} (2020) A9, [\href{http://arxiv.org/abs/1905.05697}{{\tt
  arXiv:1905.05697}}].

\bibitem{Meerburg:2012id}
P.~D. Meerburg and E.~Pajer, {\it {Observational Constraints on Gauge Field
  Production in Axion Inflation}},  {\em JCAP} {\bf 02} (2013) 017,
  [\href{http://arxiv.org/abs/1203.6076}{{\tt arXiv:1203.6076}}].

\bibitem{Bhattacharya:2017pws}
S.~Bhattacharya, K.~Dutta, M.~R. Gangopadhyay, and A.~Maharana, {\it
  {Confronting K\"ahler moduli inflation with CMB data}},  {\em Phys. Rev. D}
  {\bf 97} (2018), no.~12 123533, [\href{http://arxiv.org/abs/1711.04807}{{\tt
  arXiv:1711.04807}}].

\bibitem{Bhattacharya:2020gnk}
S.~Bhattacharya, K.~Dutta, M.~R. Gangopadhyay, A.~Maharana, and K.~Singh, {\it
  {Fibre Inflation and Precision CMB Data}},  {\em Phys. Rev. D} {\bf 102}
  (2020) 123531, [\href{http://arxiv.org/abs/2003.05969}{{\tt
  arXiv:2003.05969}}].

\bibitem{Bhattacharya:2022akq}
S.~Bhattacharya, K.~Dutta, M.~R. Gangopadhyay, and A.~Maharana, {\it
  {\ensuremath{\alpha}-attractor inflation: Models and predictions}},  {\em
  Phys. Rev. D} {\bf 107} (2023), no.~10 103530,
  [\href{http://arxiv.org/abs/2212.13363}{{\tt arXiv:2212.13363}}].

\end{thebibliography}\endgroup

\end{document}